\documentclass[11pt]{article}
\usepackage[margin=1in]{geometry}

\usepackage{newtxtext,newtxmath}
\usepackage{graphicx}    
\usepackage{amsmath}     
\usepackage{amsfonts}    
\usepackage{cite}        
\usepackage{booktabs}    
\usepackage{float}       
\usepackage{xcolor}      
\usepackage{multirow}    
\usepackage{caption}     
\usepackage{subcaption}  
\usepackage{lipsum}      
\usepackage{array}
\usepackage{tabularx} 
\usepackage{adjustbox}
\usepackage{wrapfig}
\usepackage{orcidlink}
\usepackage{hyperref}
\hypersetup{
    colorlinks=true,     
    linkcolor=black,     
    citecolor=green,     
    urlcolor=green,      
    filecolor=green,     
    pdfborder={0 0 0}    
}

\usepackage{fancyhdr}

\fancypagestyle{plain}{%
  \fancyhf{}
  \fancyhead[L]{\footnotesize Preprint}
  \fancyfoot[C]{\thepage}
  
}

\makeatletter
\def\footnoterule{%
  \kern-5pt
  \hrule \@width .5\columnwidth \@height .4pt 
  \kern 4.6pt
}
\makeatother

\newcommand{\figref}[1]{\hyperref[#1]{\textcolor{red}{\ref*{#1}}}}
\newcommand{\tabref}[1]{\hyperref[#1]{\textcolor{red}{\ref*{#1}}}}
\newcommand{\Eqref}[1]{\hyperref[#1]{\textcolor{red}{\ref*{#1}}}}

\title{Bridging Text and Video Generation: A Survey}

\author{
  Nilay Kumar\orcidlink{0009-0001-3104-6779}\textsuperscript{\textdagger},
  Priyansh Bhandari\orcidlink{0009-0006-6515-8046},
  G.\ Maragatham\orcidlink{0000-0003-1589-0571}\\[2pt]
  Department of Computational Intelligence\\
  SRM Institute of Science and Technology, KTR\\
  \texttt{\{nl9459, pr6479, maragatg\}@srmist.edu.in}
}

\date{}

\begin{document}

\maketitle

\begingroup
\renewcommand\thefootnote{\fnsymbol{footnote}}
\footnotetext[2]{Corresponding author}
\endgroup

\begin{abstract}
Text-to-video (T2V) generation technology holds immense potential to transform multiple domains such as education, marketing, entertainment, and assistive technologies for individuals with visual or reading comprehension challenges, by creating coherent visual content from natural language prompts. From its inception, the field has advanced from adversarial models to diffusion-based models, yielding higher-fidelity, more temporally consistent outputs. Yet multiple challenges still persist in these domains, such as alignment, long-range coherence, and computational efficiency. Addressing this evolving landscape, we present a comprehensive survey of text-to-video generative models, tracing their development from early GANs and VAEs to current hybrid Diffusion-Transformer (DiT) architectures, detailing how these models work internally, what limitations they addressed in their predecessors, and why shifts toward new architectural paradigms were necessary to overcome challenges in quality, coherence, and control. We provide a systematic account of the datasets, which the surveyed text-to-video models were trained and evaluated on, and, to support reproducibility and assess the accessibility of training such models, we detail their training configurations, including their hardware specifications, GPU counts, batch sizes, learning rates, optimizers, epochs, and other key hyperparameters. Further, we outline the evaluation metrics commonly used for evaluating such models and present their performance across standard benchmarks, while also discussing the limitations of these metrics and the emerging shift toward more holistic, perception-aligned evaluation strategies. Finally, drawing from our analysis, we outline the current open challenges and propose a few promising future directions, laying out a perspective for future researchers to explore and build upon in advancing T2V research and applications.
\end{abstract}

\noindent\textbf{Author’s Note (October 2025):} This paper surveys text-to-video generation models and benchmarks up to Q1 2025. Since its initial completion, several new models have emerged (e.g., the Veo series by Google, the Sora series by OpenAI, and further models released by Luma Labs, Runway, Kling AI, and others). An updated survey is planned for a future release to incorporate these developments. Until then, this version provides a detailed foundation covered in the rest of the paper.

\section{Introduction}
Deep learning has brought about a paradigm shift in generative modeling, enabling significant advancements across vision, language, and cross-modal domains. A particularly impactful area is text-to-image (T2I) synthesis, where models such as DALL·E~\cite{ramesh2021zeroshot}, CLIP~\cite{radford2021clip}, and Stable Diffusion~\cite{rombach2022ldm} have shown remarkable capabilities in generating high-fidelity images from natural language descriptions. Building on the success of T2I, recent research efforts have begun addressing the more complex task of text-to-video (T2V) generation, where the goal is to synthesize temporally coherent and semantically aligned video sequences from textual prompts. T2V extends beyond static visual depiction, requiring the modeling of motion dynamics, object persistence, temporal transitions, and scene continuity over time. This inherently makes the problem more demanding than T2I, both in terms of model architecture and computational overhead. Additionally, the alignment between natural language and visual content becomes significantly more intricate when extended to dynamic scenes, often involving multiple interacting objects and evolving spatial contexts. Despite these challenges, T2V presents immense potential across a range of real-world applications. In education, it can facilitate the visualization of abstract processes or physical phenomena, thereby enhancing comprehension and learner engagement. In accessibility contexts, it offers a means to convey information visually for users with reading difficulties or visual impairments. Within marketing and entertainment, T2V enables scalable production of product demonstrations, personalized media, and animated storytelling. However, several critical obstacles still impede its broader adoption. Ensuring smooth temporal consistency is vital to avoid visually jarring transitions~\cite{villegas2017hierarchical}. Semantic alignment between text and video remains a core challenge, particularly in multi-object or action-rich scenes~\cite{li2018dsae}. The task is further compounded by the high computational costs associated with video generation~\cite{tulyakov2018mocogan} and the limited availability of large-scale, high-quality text-video datasets~\cite{hong2018unsupervised}, both of which restrict scalability and generalization.

The field has introduced a wide range of models aimed at addressing the challenges of text-to-video (T2V) generation by extending techniques from text-to-image (T2I) synthesis to incorporate temporal dynamics. GODIVA~\cite{wu2021godiva}, one of the earlier models, employed autoencoders and attention mechanisms to efficiently generate short video clips. N\"UWA~\cite{wu2021nuwa} proposed a unified framework that considered not only text but also image and video modalities, albeit with significantly increased computational demands. In response to the scarcity of large-scale paired video datasets, models such as CogVideo~\cite{hong2022cogvideo} utilized extensive text-image datasets; however, aligning textual descriptions with temporally evolving visual content remained a substantial challenge. Make-A-Video~\cite{singer2022makeavideo} explored unsupervised zero-shot generation, but the resulting visual quality was often poor and lacked consistency. More recent research has focused on improving efficiency, temporal coherence, and output diversity. VideoFusion~\cite{luo2023videofusion} introduced a probabilistic framework to minimize redundancy while preserving coherent motion across frames. LatentShift~\cite{an2023latentshift} incorporated temporal modeling within the latent space of T2I models, offering significant reductions in computational complexity. FreeBloom~\cite{huang2023freebloom} integrated large language models to enhance narrative quality, although maintaining visual consistency across frames remains a persistent challenge. VideoTetris~\cite{tian2024videotetris} pushed the boundaries of multi-subject video generation through improved compositionality and identity preservation. Similarly, models like FIFO-Diffusion~\cite{kim2024fifodiffusion} and Pyramidal Flow Matching~\cite{jin2024pyramidal} proposed hierarchical generation strategies for long-form videos, aiming to balance high-resolution quality with temporal continuity. Evaluating the performance of T2V models continues to be a complex issue. While quantitative metrics offer scalable benchmarking, they often fail to capture nuanced aspects of visual realism, semantic alignment, and temporal coherence. In contrast, systematic human evaluation provides deeper insight into these subjective qualities, particularly in assessing alignment with textual prompts. Establishing standard evaluation protocols, combining both human and metric-based methods, will be essential to ensure fair comparisons and consistent progress across the field. Training T2V models also demands substantial computational resources and high-quality, well-curated datasets. This paper analyzes the commonly used datasets, evaluating their scale, diversity, and suitability for supporting realistic and expressive video generation. The availability of richly annotated, diverse datasets remains a critical factor in achieving high-quality T2V outputs. Despite rapid advancements, the field still lacks a comprehensive and unified overview that brings together model architectures, evaluation strategies, training protocols, and data resources. Existing surveys~\cite{singh2023survey,sun2024sora} address individual components but fall short of delivering an integrated, in‑depth review. This survey addresses that gap by systematically analyzing state-of-the-art models, evaluation methods, benchmark datasets, and training configurations. The main contributions of this paper include:
\begin{itemize}
    \item A detailed overview of the development of T2V models, covering their core architectures, training methodologies, and design principles. The paper outlines the progression from early autoencoder-based approaches to recent diffusion-based frameworks, highlighting key innovations and their impact on the field.
    
    \item A comprehensive review of the datasets commonly used for T2V model training, focusing on their scale, diversity, and content characteristics. The discussion also includes typical training setups and computational requirements, emphasizing the infrastructure necessary for effective model development.
    
    \item An analysis of the evaluation metrics employed in the T2V domain, with attention to current benchmarking practices. The paper also examines the role of human evaluations in assessing realism and semantic alignment, and reviews emerging protocols aimed at improving evaluation reliability.
    
    \item A discussion of the central challenges facing T2V research, including limited data availability, high computational costs, and difficulties in modeling temporal consistency. The paper outlines potential solutions, such as synthetic data generation, architectural adjustments, and improved temporal modeling, while emphasizing the need to broaden real-world applications.
\end{itemize}

The rest of the paper is organized as follows: Section~\ref{sec:preliminaries} covers the preliminaries. Section~\ref{sec:methodologies} presents the methodologies. Section~\ref{sec:datasets} details the datasets and training configurations. Section~\ref{sec:evaluation} outlines the evaluation metrics and benchmarks. Section~\ref{sec:future} discusses key challenges and future directions, and finally, Section~\ref{sec:conclusion} concludes the paper. 

\section{Preliminaries}
\label{sec:preliminaries}

The development of text-to-video (T2V) models has been deeply shaped by progress in foundational architectures originally designed for image synthesis and sequence modeling. To understand how T2V systems generate coherent videos from text prompts, it is important to first establish the key architectural and learning paradigms that underpin this field. These include encoder–decoder frameworks for spatial understanding, probabilistic methods for latent representation learning, adversarial strategies for realism, diffusion-based sampling for high-fidelity generation, and attention mechanisms for capturing long-range dependencies across modalities and time. Each of these foundations has played a pivotal role in enabling the transition from static image generation to dynamic, temporally consistent video synthesis. In the subsections that follow, we briefly introduce these core components to provide the necessary context for understanding the evolution and design of modern T2V models.

\subsection{U-Net}

U-Net~\cite{ronneberger2015unet} belongs to the class of convolutional neural networks~\cite{lecun1989backprop} that were specifically designed for dense prediction tasks that require both global context and local detail preservation. U-Net is built up of two core components: (1) an Encoder and (2) a Decoder, that work together to transform input features into detailed output maps. In the U-Net framework, the network architecture is symmetric in the shape of a ``U''. The network is built of an encoder module, which represents the contracting path, and a decoder module, which represents the expanding path. In this approach, the encoder gathers contextual information by continuously decreasing the input image's spatial scale and simultaneously enhancing the richness of feature representations through subsequent layers of convolution, followed by pooling operations, as shown in Eqs.~(\Eqref{eq:conv})--(\Eqref{eq:pool}):

\begin{align}
f_l &= \sigma(W_l * f_{l-1} + b_l) \label{eq:conv} \\
f_l &= \mathit{Pool}(f_l) \label{eq:pool}
\end{align}

\noindent where $\sigma(\cdot)$ is a nonlinear activation function like ReLU~\cite{glorot2011deep}, $W_l$ and $b_l$ are the weights and biases, and $\mathit{Pool}$ typically refers to max pooling~\cite{lecun1989backprop}. The decoder reconstructs the spatial dimensions and refines feature maps for precise localization. Upsampling is performed by concatenating the upsampled features with corresponding encoder features via skip connections~\cite{he2016resnet}, followed by convolutional operations, as shown in Eqs.~(\Eqref{eq:up})--(\Eqref{eq:decode}):

\begin{align}
f_l' &= \mathit{Up}(f_{l+1})                             \label{eq:up} \\
f_l^{\mathit{concat}} &= [\,f_l',\, f_{E_l}\,]           \label{eq:concat} \\
f_l &= \sigma\!\big(W_l' * f_l^{\mathit{concat}} + b_l'\big) \label{eq:decode}
\end{align}

\noindent where $\mathit{Up}(\cdot)$ denotes upsampling (e.g., transposed convolution~\cite{zeiler2010deconvnet}), and $f_{E_l}$ are the encoder feature maps at level~$l$. U-Net learns a mapping from an input image $\mathit{x} \in \mathit{\mathbb{R}}^{\mathit{H} \times \mathit{W} \times \mathit{C}}$ to an output segmentation map $\mathit{y} \in \mathit{\mathbb{R}}^{\mathit{H} \times \mathit{W} \times \mathit{K}}$ by minimizing the cross-entropy loss, denoted in Eq.~(\Eqref{eq:unetloss}):

\begin{equation}
\mathit{\mathcal{L}}(\mathit{\theta}; \mathit{x}, \mathit{y}) = - \sum_{i=1}^{\mathit{H} \times \mathit{W}} \sum_{k=1}^{\mathit{K}} \mathit{y}_{i,k} \log \mathit{p}_{i,k}
\label{eq:unetloss}
\end{equation}

where $p_{i,k}$ is the predicted probability for pixel $i$ and class $k$. By minimizing this loss, U-Net effectively captures abstract semantic information through its encoder and retains detailed spatial features via its decoder and skip connections. This dual capability makes the U-Net an extremely strong architecture for tasks that call for both global understanding and pixel-level precision, extending its application from segmentation into a wide array of image transformation and generation tasks as demonstrated in~\cite{isola2017pix2pix, ledig2017srgan, pathak2016contextencoders, johnson2016perceptual, zhu2017cyclegan}.

\subsection{Variational Auto-Encoders (VAEs)}

Variational Auto-Encoders (VAEs)~\cite{kingma2013vae} are a class of deep generative models built on the principles of Bayes' theorem~\cite{bayes1763essay}, aiming to learn continuous latent representations of data. A VAE is composed of two neural networks—an encoder and a decoder—that collaboratively model complex data distributions. In the VAE framework, the goal is to model an unknown data distribution $\mathit{p}(\mathit{x})$ by introducing latent variables $\mathit{z}$ that capture the underlying structure of the data. The encoder network $\mathit{q}_\mathit{\phi}(\mathit{z}|\mathit{x})$, parameterized by $\mathit{\phi}$, approximates the intractable posterior over latent variables conditioned on the observed data. It encodes the input $\mathit{x}$ into a latent distribution, effectively capturing salient features in a compressed form. The decoder network $\mathit{p}_\mathit{\theta}(\mathit{x}|\mathit{z})$, parameterized by $\mathit{\theta}$, models the likelihood of the data conditioned on latent variables and reconstructs the input from its encoding. The joint probability of the observed data $\mathit{x}$ and the latent variables $\mathit{z}$ is defined in Eq.~(\Eqref{eq:joint}):

\begin{equation}
\mathit{p}_\mathit{\theta}(\mathit{x}, \mathit{z}) = \mathit{p}_\mathit{\theta}(\mathit{x}|\mathit{z}) \, \mathit{p}(\mathit{z})
\label{eq:joint}
\end{equation}

Here, $\mathit{p}(\mathit{z})$ denotes the prior over the latent space, usually assumed to follow a standard normal distribution $\mathcal{N}(0, \mathit{I})$. The term $\mathit{p}_\mathit{\theta}(\mathit{x}|\mathit{z})$ defines the generative process through which the decoder reconstructs the input data. Training a VAE involves maximizing the Evidence Lower Bound (ELBO)~\cite{bishop1995vpc}, a variational objective that serves as a proxy for the marginal log-likelihood of the observed data, defined in Eq.~(\Eqref{eq:elbo}):


\begin{equation}
\label{eq:elbo}
\adjustbox{max width=\linewidth, valign=c}{$
\mathit{L}(\mathit{\theta},\mathit{\phi};\mathit{x})
= \mathbb{E}_{\mathit{q}_{\mathit{\phi}}(\mathit{z}\mid\mathit{x})}\big[\log \mathit{p}_{\mathit{\theta}}(\mathit{x}\mid\mathit{z})\big]
- \mathit{KL}\!\left(\mathit{q}_{\mathit{\phi}}(\mathit{z}\mid\mathit{x})\,\|\,\mathit{p}(\mathit{z})\right)
$}
\end{equation}

The first term, $\mathbb{E}_{\mathit{q}_{\mathit{\phi}}(\mathit{z} \,|\, \mathit{x})}[\log \mathit{p}_{\mathit{\theta}}(\mathit{x} \,|\, \mathit{z})]$, represents the expected reconstruction accuracy, guiding the decoder to produce outputs that resemble the original input $\mathit{x}$. The second term, $\mathit{KL}(\mathit{q}_{\mathit{\phi}}(\mathit{z} \,|\, \mathit{x}) \, \| \, \mathit{p}(\mathit{z}))$, is the Kullback-Leibler divergence, acting as a regularizer that aligns the learned posterior with the prior and encourages smoothness in the latent space. By optimizing the ELBO with respect to both $\mathit{\phi}$ and $\mathit{\theta}$, VAEs balance reconstruction quality and latent space regularity. Once trained, they can generate new samples by sampling from the prior $\mathit{p}(\mathit{z})$ and decoding these samples through $\mathit{p}_{\mathit{\theta}}(\mathit{x} \,|\, \mathit{z})$, thereby capturing the essential structure of the original data distribution.

\subsection{Generative Adversarial Networks (GANs)}

GANs~\cite{goodfellow2014gan} are a type of deep generative model that frames data creation as an interplay between two neural nets: the neural nets being the (1) Generator and the (2) Discriminator. These networks engage in a competitive process, with each network refining its ability to generate or distinguish data. GANs utilize adversarial training to learn complex data distributions without explicitly modelling probability density functions. The generator $\mathit{G}_{\mathit{\theta}}(\mathit{z})$, parameterized by $\mathit{\theta}$, transforms random noise vectors $\mathit{z}$ from a prior distribution $\mathit{p}_{\mathit{z}}(\mathit{z})$ (typically a standard normal distribution) into synthetic data samples $\mathit{x}$. The discriminator $\mathit{D}_{\mathit{\phi}}(\mathit{x})$, parameterized by $\mathit{\phi}$, evaluates whether a given sample $\mathit{x}$ is real (from $\mathit{p}_{\text{data}}(\mathit{x})$) or fake (produced by $\mathit{G}_{\mathit{\theta}}$). The GAN framework aims to have the generator's distribution $\mathit{p}_{\mathit{g}}(\mathit{x})$ closely approximate the true data distribution $\mathit{p}_{\text{data}}(\mathit{x})$. This is achieved through a minimax optimization objective, denoted in Eq.~(\Eqref{eq:min_max}):


\begin{equation}
\label{eq:min_max}
\adjustbox{max width=\linewidth, valign=c}{$
\mathcal{L}_{\text{adv}}(\theta,\phi)
= \min_{\theta}\max_{\phi}\,\big[
\mathbb{E}_{x \sim p_{\text{data}}(x)} \log D_{\phi}(x)
+ \mathbb{E}_{z \sim p_z(z)} \log\!\big(1 - D_{\phi}(G_{\theta}(z))\big)
\big]
$}
\end{equation}

In this setup, the discriminator $\mathit{D}_\mathit{\phi}$ gets better in maximizing its accuracy in distinguishing real from fake samples, while the generator $\mathit{G}_\mathit{\theta}$ strives to minimize the discriminator's ability to make correct classifications. Training involves iteratively updating both networks: the discriminator maximizes its objective $\mathit{L}_\mathit{D}(\mathit{\phi})$, and the generator minimizes its objective $\mathit{L}_\mathit{G}(\mathit{\theta})$. However, to improve gradient flow and enhance training stability, the generator's loss is often redefined as in Eq.~(\Eqref{eq:genloss}):

\begin{equation}
\mathit{L}_\mathit{G}'(\mathit{\theta}) =
- \mathbb{E}_{\mathit{z} \sim \mathit{p}_\mathit{z}(\mathit{z})}
\left[ \log \mathit{D}_\mathit{\phi}\!\left( \mathit{G}_\mathit{\theta}(\mathit{z}) \right) \right]
\label{eq:genloss}
\end{equation}

By repetitively optimizing this objective, GANs learn to generate realistic data samples that can hardly be distinguished from the real data. While the adversarial training allows the generator to capture highly intricate data distributions, the discriminator continuously adapts to detect extremely subtle differences between generated and real data. This has opened up numerous use cases in the fields of computer vision and image synthesis like image generation~\cite{goodfellow2014gan}, style transfer~\cite{johnson2016perceptual}, image-super-resolution~\cite{ledig2017srgan}, powered by the strong generative capabilities of GANs.

\subsection{Denoising Diffusion Probabilistic Models (DDPMs)}

DDPMs~\cite{ho2020ddpm} are a sophisticated family of deep learning based generative models that generate data using progressive denoising, drawing inspiration from non-equilibrium thermodynamics. DDPM functions in two stages: (\textbf{a}) a forward diffusion phase where Gaussian noise is gradually introduced to the input, and (\textbf{b}) a backward phase that aims to reconstruct the original input by incrementally removing the noise. It leverages the simplicity of the Gaussian distribution together with the representation power of neural networks to model complex distributions without explicit likelihood estimation.

\subsubsection{Unconditioned DDPMs}
In unconditioned DDPMs, the forward diffusion process gradually converts an initial data point into a sequence of noisy latent variables by adding Gaussian noise at each timestep. This process is regulated by a variance schedule, where each value in the schedule determines the noise intensity at a specific timestep, denoted in Eq.~(\Eqref{eq:ddpm_forward}):

\begin{equation}
\mathit{q}(x_t | x_{t-1}) = \mathcal{N}(x_t; \sqrt{1 - \beta_t} \cdot x_{t-1}, \beta_t \mathit{I})
\label{eq:ddpm_forward}
\end{equation}

Here, $\mathit{q}(x_t | x_{t-1})$ denotes the probability distribution of the noisy variable at timestep $t$ given $t{-}1$; $\mathcal{N}$ is a Gaussian distribution; $x_t$ and $x_{t-1}$ are the noisy latent variables; $\beta_t$ controls the noise added; and $\mathit{I}$ is the identity matrix for uniform noise application. The aggregate spread at any timestep $t$ is defined as in Eq.~(\Eqref{eq:ddpm_marginal}):

\begin{equation}
\mathit{q}(x_t | x_0) = \mathcal{N}(x_t; \sqrt{\bar{\alpha}_t} \cdot x_0, (1 - \bar{\alpha}_t)\mathit{I})
\label{eq:ddpm_marginal}
\end{equation}

Here, $\mathit{q}(x_t | x_0)$ represents the aggregate spread of $x_t$ at timestep $t$ given $x_0$, where $\alpha_t = 1 - \beta_t$ and $\bar{\alpha}_t = \prod_{s=1}^{t} \alpha_s$, which control the noise schedule. The reverse denoising process reconstructs the initial data point from a sequence of noisy variables, as defined in Eq.~(\Eqref{eq:ddpm_reverse}):

\begin{equation}
\mathit{p}_\theta(x_{t-1} | x_t) = \mathcal{N}(x_{t-1}; \mu_\theta(x_t, t), \sigma_t^2 \mathit{I})
\label{eq:ddpm_reverse}
\end{equation}

Here, $\mathit{p}_\theta(x_{t-1} | x_t)$ models the reverse denoising to estimate $x_{t-1}$ from $x_t$. $\mu_\theta(x_t, t)$ is a mean predicted by a neural network, and $\sigma_t^2$ is a variance, either fixed or learnable, that adjusts noise reduction at each step. Training involves minimizing the loss function, denoted in Eq.~(\Eqref{eq:ddpm_loss}):

\begin{equation}
\mathcal{L}_{\text{simple}}(\mathit{\theta}) = \mathbb{E}_{\mathit{x}_0, \mathit{\epsilon}, \mathit{t}} \left[\| \mathit{\epsilon} - \mathit{\epsilon}_\theta(\mathit{x}_t, \mathit{t}) \|^2\right]
\label{eq:ddpm_loss}
\end{equation}

where $\mathit{x}_t = \sqrt{\bar{\alpha}_t} \cdot \mathit{x}_0 + \sqrt{1 - \bar{\alpha}_t} \cdot \mathit{\epsilon}$ and $\mathit{\epsilon} \sim \mathcal{N}(0, \mathit{I})$. This objective trains the model to accurately predict and remove the added noise, enabling the generation of high-fidelity samples from pure noise.

\subsubsection{Conditioned DDPMs}
Conditioned DDPMs extend the unconditioned framework by incorporating additional information to guide the generation process, enabling the creation of data that aligns with specific attributes or inputs. This conditioning is achieved through two primary methods: classifier guidance~\cite{dhariwal2021diffusion} and classifier-free guidance~\cite{ho2022classifierfree}.

\textbf{(i) Classifier Guidance}:
In Classifier Guidance, the model estimates the likelihood of a target attribute based on noisy inputs. During the reverse diffusion process, it uses gradient information to refine predictions, as shown in Eq.~(\Eqref{eq:classifier_guidance}):

\begin{equation}
\label{eq:classifier_guidance}
\adjustbox{max width=\linewidth, valign=c}{$
p_\theta(x_{t-1}\,|\,x_t,y)
= \mathcal{N}\!\left(
x_{t-1}\,;\,
\mu_\theta(x_t,t) + s\,\sigma_t^2 \nabla_{x_t}\log p_\phi(y\,|\,x_t)\,,\,
\sigma_t^2 I
\right)
$}
\end{equation}

\noindent where $\mathit{p}_\phi(\mathit{y} \,|\, \mathit{x}_t)$ serves as an external classifier to guide the prediction, $\mathit{s}$ adjusts the strength of guidance, and the gradient $\nabla_{\mathit{x}_t} \log \mathit{p}_\phi(\mathit{y} \,|\, \mathit{x}_t)$ steers sampling toward regions aligned with the target attribute $\mathit{y}$.

\textbf{(ii) Classifier-Free Guidance}:
Classifier-free guidance integrates conditioning directly into the denoising model, removing the need for an external classifier. The model $\mathit{\epsilon}_\theta(\mathit{x}_t, \mathit{t}, \mathit{y})$ is trained to predict noise with and without conditioning information by randomly omitting $\mathit{y}$ during training. At inference, the prediction is adjusted as in Eq.~(\Eqref{eq:cfg}):

\begin{equation}
\hat{\epsilon}_\theta(x_t, t \,|\, y)
= (1 + w)\,\epsilon_\theta(x_t, t \,|\, y) \;-\; w\,\epsilon_\theta(x_t, t)
\label{eq:cfg}
\end{equation}

Here $w$ denotes the guidance scale, which fine-tunes the balance between sample diversity and fidelity to the conditioning information. Both guidance methods enhance DDPMs' ability to generate conditioned data, enabling applications like conditional image synthesis and text-to-image generation~\cite{rombach2022ldm} by effectively steering the denoising process toward desired attributes.

\subsection{Transformers}

Transformers~\cite{vaswani2017attention} represent a revolutionary class of deep models that have brought a sea change in NLP and now find growing applications in computer vision and studies on generation~\cite{vaswani2017attention, dosovitskiy2020vit, radford2019gpt2}. Transformer models utilize mechanisms of self-attention so as to model dependencies between input and output sequences. This enables the parallel computation of the input and allows for capturing long-range interactions efficiently. This architecture is made up of two major modules: Firstly, The Encoder and Secondly, The Decoder, both of which include multiple layers; commonly these are made up of self-attention modules and feed-forward neural network layers. The encoder is designed to identify contextual relationships using self-attention, whereas the decoder is tasked with creating output sequences based on attention to the encoder representations and previously generated outputs. At the core of Transformers is the self-attention mechanism, defined as in Eq.~(\Eqref{eq:attention}):

\begin{equation}
\mathit{Attention}(\mathit{Q}, \mathit{K}, \mathit{V}) = \mathit{softmax}\left( \frac{\mathit{Q} \mathit{K}^\top}{\sqrt{\mathit{d}_k}} \right) \mathit{V}
\label{eq:attention}
\end{equation}

The formula in Eq.~\eqref{eq:attention} defines how the attention mechanism computes the interaction between query ($\mathit{Q}$), key ($\mathit{K}$), and value ($\mathit{V}$) matrices, which originate from input embeddings. Here, $\mathit{d}_k$ represents the size of the key vectors. This module enables the model to dynamically assess and prioritize different elements in a sequence. The multi-head attention mechanism improves model performance by distributing inputs across various subspaces, which helps the model identify a wide range of relationships. The formula for this is shown in Eqs.~(\Eqref{eq:multihead})--(\Eqref{eq:headn}).

\begin{equation}
\mathit{MultiHead}(\mathit{Q}, \mathit{K}, \mathit{V}) = \mathit{Concat}(\mathit{head}_1, \dots, \mathit{head}_h) \, \mathit{W}^O
\label{eq:multihead}
\end{equation}

\begin{equation}
\mathit{head}_n = \mathit{Attention}(\mathit{Q} \mathit{W}_n^Q, \mathit{K} \mathit{W}_n^K, \mathit{V} \mathit{W}_n^V)
\label{eq:headn}
\end{equation}

Here, $\mathit{W}_n^Q$ is the projection matrix for queries, $\mathit{W}_n^K$ for keys, and $\mathit{W}_n^V$ for values, transforming the respective inputs to align with specific attention functions. $\mathit{W}^O$ is the final projection matrix that combines all attention heads' outputs into a single output vector. Training Transformers involves minimizing the cross-entropy loss for sequence generation, denoted in Eq.~(\Eqref{eq:transformerloss}):

\begin{equation}
\mathcal{L}(\mathit{\theta}) = -\sum_{t=1}^{T} \log P(\mathit{y}_t \mid \mathit{y}_{<t}, \mathit{x})
\label{eq:transformerloss}
\end{equation}

In Eq.~(\Eqref{eq:transformerloss}), $\mathit{y}_t$ represents the token generated at position $t$, $\mathit{y}_{<t}$ denotes the tokens that come before it, and $\mathit{x}$ refers to the original series of input data. By refining this loss function, Transformers can efficiently grasp complex dependencies and produce outputs of superior quality. These models excel at interpreting the nuanced meanings of input data, resulting in well-organized outputs, as a result, making Transformers essentially indispensable for a lot of the sequence-to-sequence applications like translation~\cite{vaswani2017attention}, summarization~\cite{liu2019textsum}, and generative modelling~\cite{dosovitskiy2020vit}. It’s with this flexibility and scalability, that Transformers have become the foundational architectures in modern deep learning applications.

\section{Methodologies}
\label{sec:methodologies}

Various deep learning approaches are contributing remarkably towards in-text-to-video generation with each solving unique aspects of synthesizing temporally coherent videos from textual descriptions. Broadly, a host of such approaches can be put under the umbrella of three such model architecture classes. Ordered by the evolution of such model classes are (a) GANs, then (b) VAEs and finally (c) DDPMs. GAN-based approaches were one of the first to be applied to generating text-to-video problems, relying on an adversarial training process between its generator and discriminator networks in order to create realistic video content. Although such models showed early video synthesis success, they are generally plagued with issues of stability during training and scaling to higher resolutions. Another point of view was introduced by VAEs-based models: they learn compact latent representation of videos that enables more controlled generation by probabilistic modeling. Such models appeared to be more stable than GANs but sometimes struggled with the quality of their output and fine details. More recently, the dominant paradigm has been the emergence of diffusion-based models for the synthesis of text-based video generation that exhibit superior quality and have more temporal consistency. These models— which have extended successful results of diffusion-based approaches for image generation— have achieved high-quality video generations with semantic alignments to input text. The overriding trend for using diffusion models is captured in many recent developments in this class of methods, as will be seen by the lengthy list of methods in this survey based on diffusion. Each of the paradigms enjoys different advantages and issues while solving the inherent problems of video generation, namely temporal consistency, visual quality, and text-video alignment. In the following sections, a thorough review of representative models from each category is provided, including discussions of architecture, methodology, and contribution.

\subsection{GANs Based Models}

GANs were introduced in 2014 by Goodfellow et al. in~\cite{goodfellow2014gan}, immediately causing a revolution in generative modeling by introducing the adversarial training paradigm. It is made up of two competing neural networks: one, the generator, generates data samples; the other, the discriminator, distinguishes between real and generated samples. When they were first designed for generating images, GANs immediately started working art wonders, managing to generate several high-quality realistic images. This immediately called for significant changes in the architecture of GANs, including temporal dynamics and consistency between frames. The different approaches and novelties authors have developed to extend GANs for video generation are discussed as follows.

\subsubsection{MoCoGAN}

\begin{figure}[htbp]
    \centering
    \includegraphics[width=0.65\columnwidth]{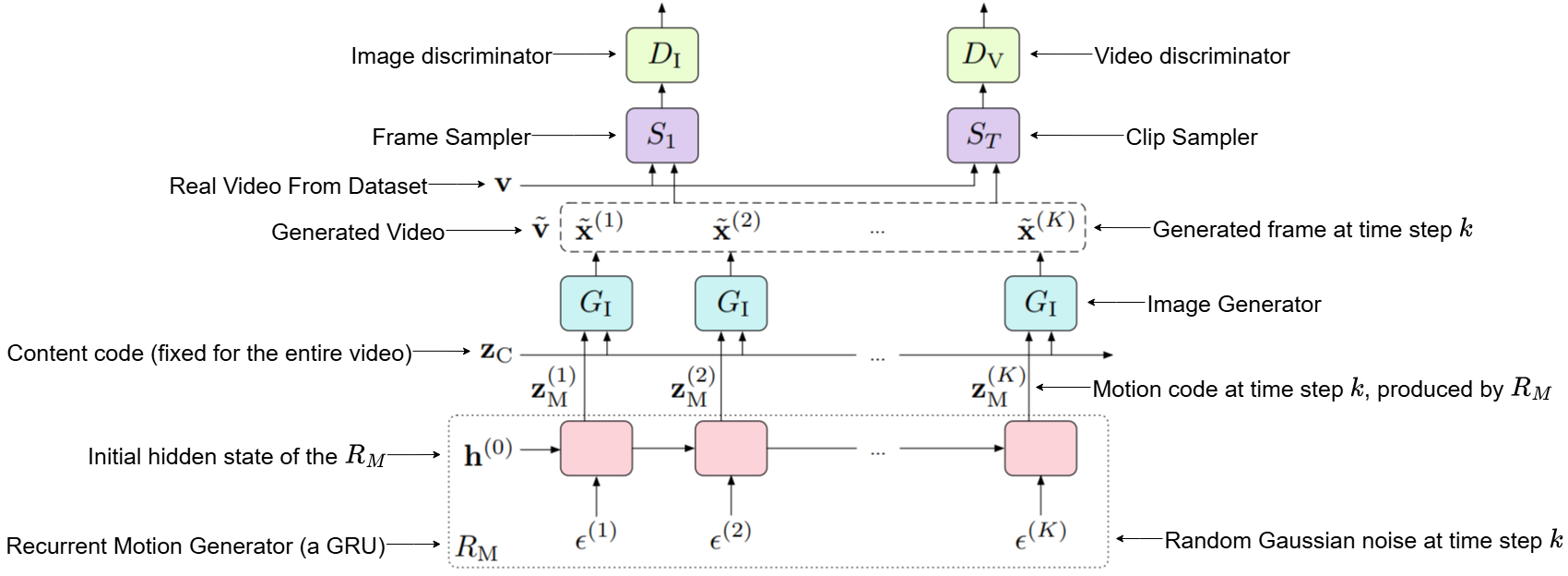}
    \caption{Model Architecture of MoCoGAN from~\cite{tulyakov2018mocogan}.}
    \label{fig:mocogan}
\end{figure}

MoCoGAN~\cite{tulyakov2018mocogan}, as illustrated in Fig.~\figref{fig:mocogan}, makes use of a decomposed GAN architecture for video generation where the latent space gets decomposed into a content subspace and a motion subspace. Such separation enables the model to handle static elements, i.e., content, and dynamic variations, i.e., motion, separately. This is controlled by a single content vector, sampled from a Gaussian distribution and kept constant throughout the video, encoding features that remain consistent across frames, like the object identity or background. The changes with respect to time are described by the motion subspace through the generation, on each frame, of motion vectors via the RNN~\cite{rumelhart1986backprop} with a Gated Recurrent Unit~\cite{cho2014rnn}, introducing a recurrent path to be able to emulate smooth, sequential motion. MoCoGAN treats the video sequences as trajectories in a latent space, where each point in that latent space takes a frame, and variations in path length allow for videos of variable duration. It contains one generator that composites a single fixed content vector with frame-specific motion vectors and two discriminators: an image discriminator for ensuring coherence in individual frames and a video discriminator for temporal coherence across sequences of frames. The discriminator in the video also includes categorical dynamics through predefined motion categories that can generate MoCoGAN with specific actions—such as walking and jumping—input, given a categorical variable.

\subsubsection{N\"UWA}

\begin{figure}[htbp]
    \centering
    \includegraphics[width=0.65\columnwidth]{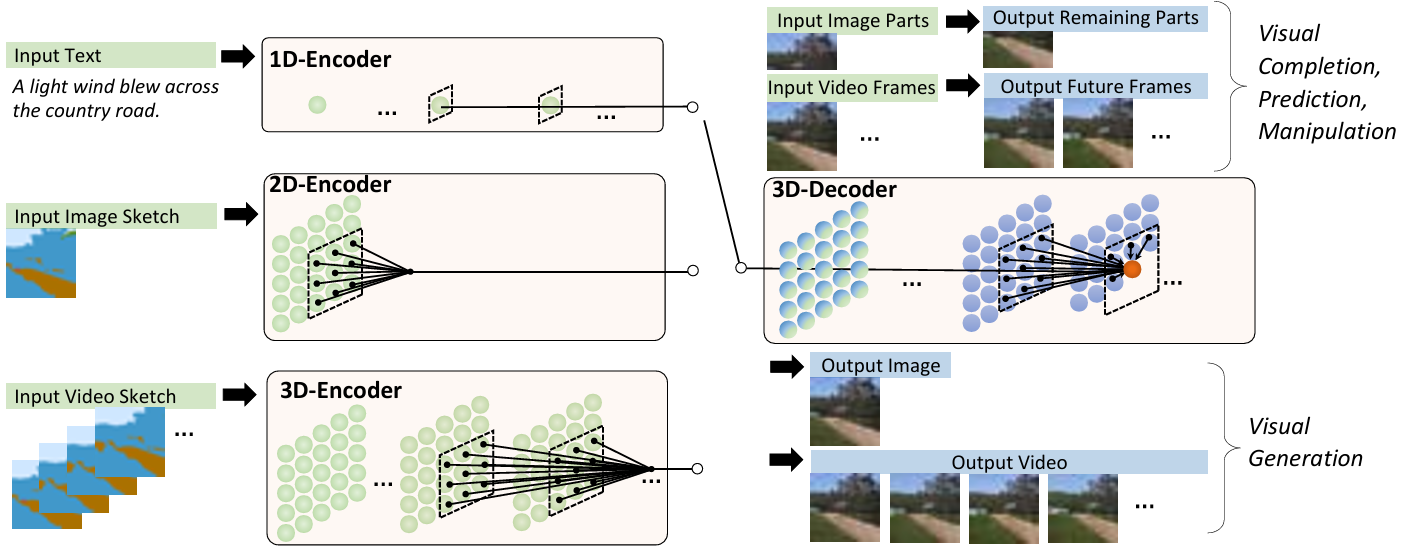}
    \caption{Model Architecture of N\"UWA from~\cite{wu2021nuwa}.}
    \label{fig:nuwa}
\end{figure}

N\"UWA presented in~\cite{wu2021nuwa} has a unified architecture of a 3D Transformer encoder-decoder, which leverages tokenization from the 2D VQ-GAN~\cite{esser2020taming} and uses 3D Nearby Self-Attention (3DNA) for holistic visual synthesis across images, videos, and textual descriptions, displayed in Fig.~\figref{fig:nuwa}. Images and videos are represented as a 3D spatial-temporal cube where each token in the cube corresponds to a patch of the visual input. Synchronized by VQ-GAN, this is a patch-based approach that tokenizes frames into discrete latent codes mapped to a shared codebook, hence allowing a unified representation of both images and videos but decreased data complexity while preserving essential information in space and time. The 3D Nearby Self-Attention mechanism in N\"UWA is focused on enhancing efficiency by considering only the spatial and temporal neighborhoods—that is, where each token attends to only the tokens in the near neighborhood inside the cube. In addition, the attention model in this work conveys local dependencies while considering coherency across frames as one of the majors for generating realistic visuals for long sequences. Training of the model is incorporated with multi-task objectives for supporting tasks like video prediction, text-to-image, and text-to-video generation where conditioning generation is either based on the conditioning text or previous frames through masked token prediction within the 3D spatial-temporal cube. This allows the model to create the missing components by capturing contextual dependencies within the spatial domain as well as in the temporal dimension.

\subsection{VAEs Based Models}
VAEs, First proposed in~\cite{kingma2013vae}, are a suite of deep generative models which try to learn to compress data to a lower latent space and later try to rebuild it back as similar as it can to the original data, trying to maintain the most important features. Unlike GANs, VAEs provide a probabilistic framework for generation and explicitly model the data distribution through variational inference. Their ability to learn structured latent representations made them particularly attractive for complex generative tasks. The next sections discuss how researchers have adapted and extended the VAE framework to video generation tasks by incorporating temporally modelling and text conditioning mechanisms.

\subsubsection{VideoGPT}

\begin{figure}[htbp]
    \centering
    \includegraphics[width=0.65\columnwidth]{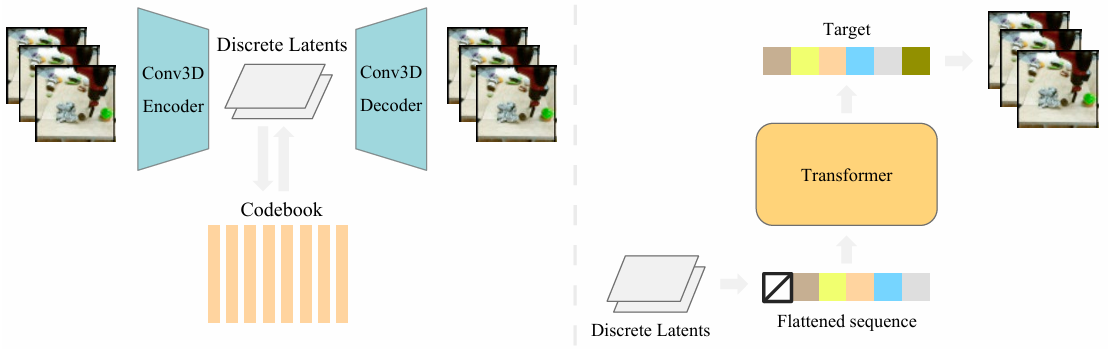}
    \caption{Model Architecture of VideoGPT from~\cite{yan2021videogpt}.}
    \label{fig:videogpt}
\end{figure}

VideoGPT~\cite{yan2021videogpt} implements a hybrid architecture, which merges VQ-VAE~\cite{oord2017vqvae} with GPT-style transformers~\cite{radford2018gpt} for efficient naturalistic video generation. It makes use of 3D convolutions~\cite{tran2014c3d} in the VQ-VAE framework for extracting deep spatial and temporal features from the videos and compressing them into a discrete latent space. Such compression reduces the data dimensionality extensively but preserves the most salient information that is critical for maintaining quality in generated videos. These 3D convolutions are used by the encoder component to transform the high-dimensional video data into lower-dimension and discrete latent representation. The decoder reconstructs video from this compressed format using transposed 3D convolutions that guarantee detailed and accurate visual outputs. The model treats the latent space by first discretizing raw video into discrete tokens through some process of vector quantization, as shown in Fig.~\figref{fig:videogpt}. The encoder outputs get matched to a learned codebook to reduce the video data into simple, manipulable codes. This projects such dense video representations into a finite set of discrete codes; therefore, it summarizes video content in a very effective way. The integration then involves a GPT-style autoregressive transformer model which predicts these latent codes sequentially with the goal of generating video frames that are coherent temporally. This involves learned spatio-temporal position encodings~\cite{radford2018gpt} to help maintain continuity both of space and time between frames.

\subsubsection{GODIVA}

\begin{figure}[htbp]
    \centering
    \includegraphics[width=0.65\columnwidth]{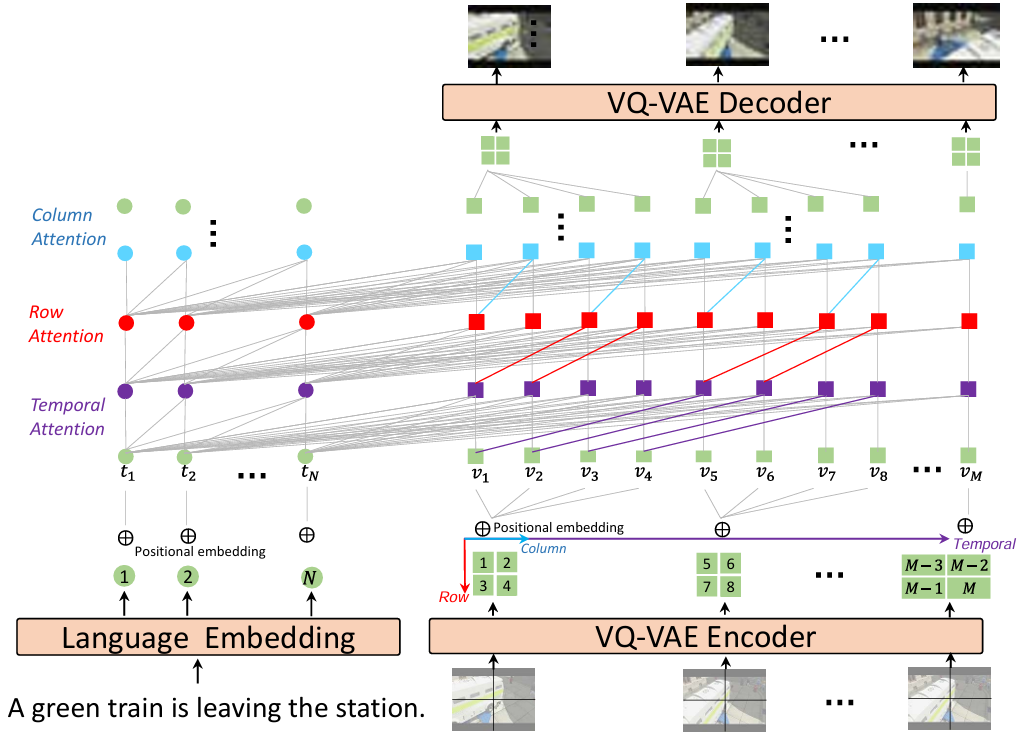}
    \caption{Model Architecture of GODIVA from~\cite{wu2021godiva}.}
    \label{fig:godiva}
\end{figure}

GODIVA, introduced in~\cite{wu2021godiva}, integrates a frame-wise VQ-VAE with 3D sparse attention~\cite{zhang2022detr4d} for video generation conditioned on natural language descriptions, as shown in Fig.~\figref{fig:godiva}. In the model, the internal VQ-VAE framework processes frames independently: each frame is encoded into latent variables via 3D convolutions accurately quantized by matching encoded regions against a codebook. This results in a discrete and compact latent representation, which is reconstructed by the decoder through transposed 3D convolutions~\cite{dosovitskiy2015checkerboard} with the preservation of the critical spatial and temporal information. VQ-VAE's training objective in GODIVA contains loss functions such as reconstruction loss, codebook loss, and commitment loss to ensure proper frame encoding and decoding. GODIVA generates video using text based on the modeled conditional probability \(P(z \mid t)\), where \(z\) stands for the discrete latent code, and \(t\) stands for the input text. Text embeddings combine the outputs from pre-trained embeddings with positional encodings to temporally align text and video. GODIVA processes the temporal, row, and column information of the encoded frames separately with its three-dimensional sparse attention mechanism that optimizes computational efficiency by attending only to the relevant spatio-temporal features.

\subsubsection{CogVideo}

\begin{figure}[htbp]
    \centering
    \includegraphics[width=0.65\columnwidth]{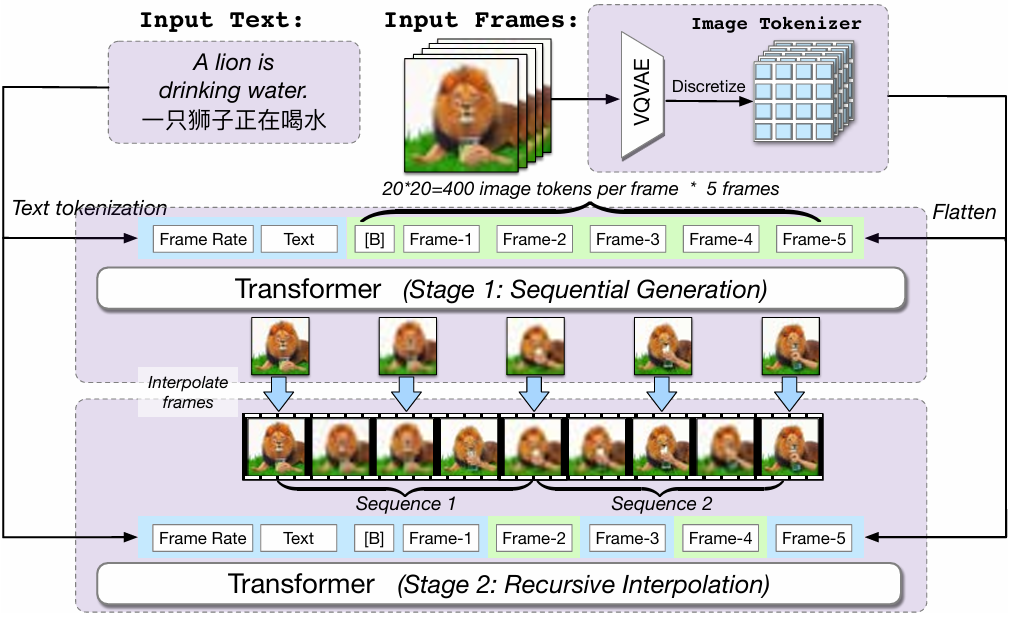}
    \caption{Model Architecture of CogVideo from~\cite{hong2022cogvideo}.}
    \label{fig:cogvideo}
\end{figure}

CogVideo~\cite{hong2022cogvideo} implements a dual-channel transformer-based architecture for large-scale text-to-video generation, as shown in Fig.~\figref{fig:cogvideo}. Within its framework, the spatial and temporal information is separately treated by the dual-channel attention blocks. It contains two major types of attention channels in its architecture: an attention-base layer pre-trained for spatial features and an attention-plus layer designed for temporal processing. Further, these channels dynamically combine through a learnable mixture factor, which effectively weights the adaptation between spatial and temporal components. This transformer-based architecture effectively generates aligned videos through the incorporation of several multi-modal attention mechanisms. The generation of the video pipeline, designed to work in a multi-stage process, first goes through a sequential generation phase at low frame rates, creating keyframes, and is followed by a recursive interpolation phase to eventually create intermediate frames for temporal coherence. Its architecture features two important components, including the 3D Shifted Window attention~\cite{li2023r3dswin} that earmarks the efficient processing of the spatial-temporal cube by applying localized attention to neighboring spatial and temporal tokens. This rewindowing strategy minimizes memory usage while allowing parallelization of spatial and temporal dimensions. The architecture includes mechanisms for conditioning on the frame rate, giving control on the output frame rate to secure alignment with textual specifications.

\subsection{Diffusion Models}
Building upon the theoretical developments of Sohl-Dickstein et al. in 2015, recently emerged as one of the successful generative models. These models work progressively to denoise random Gaussian noise in a manner guided by a learned reverse diffusion process. Their success in generating images through models like DALL·E~\cite{ramesh2021zeroshot} and Stable Diffusion~\cite{rombach2022ldm} has led to their rapid adoption for video synthesis. Of course, diffusion models boast several advantages, such as stable training dynamics and generation of high quality. In the subsequent sections, the paper describes the various approaches in relation to extending diffusion models towards video generation by considering temporal consistency, computational efficiency, and alignment between input texts and generated videos.

\subsubsection{Make-A-Video}

\begin{figure}[htbp]
    \centering
    \includegraphics[width=0.65\columnwidth]{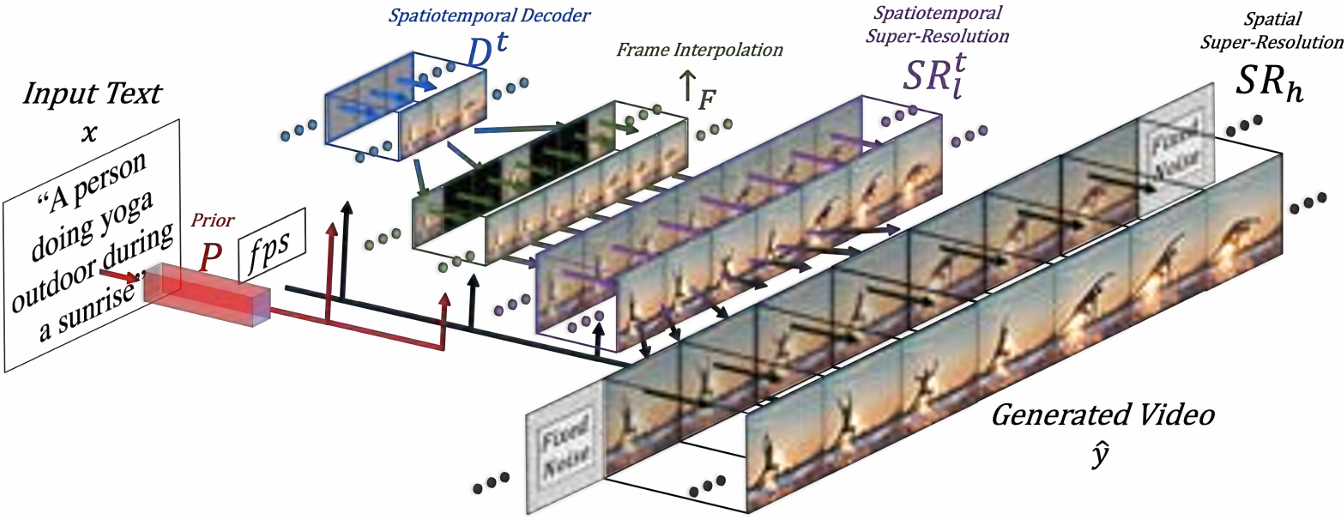}
    \caption{Model Architecture of Make-A-Video from~\cite{singer2022makeavideo}.}
    \label{fig:makeavideo}
\end{figure}

Make-A-Video~\cite{singer2022makeavideo} follows a multi-stage diffusion-based architecture that generates videos leveraging a pre-trained text-to-image (T2I) model and has no need for paired text-video datasets, as shown in Fig.~\figref{fig:makeavideo}. The architecture uses a pre-trained T2I diffusion model, spatiotemporal layers to capture temporal dynamics, and a frame interpolation network to better the continuity between frames. At the beginning of the process, the model follows a pre-trained T2I diffusion model to generate base frames at \(64 \times 64\) resolution. These frames are later enhanced by two specialized super-resolution networks: SR\textsubscript{t} is a spatiotemporal network that increases the resolution of the frames to \(256 \times 256\) and ensures temporal coherence, whereas SR\textsubscript{h} is used for spatial super-resolution, increasing the resolution further up to \(768 \times 768\). The architecture extends T2I capabilities into video synthesis through the use of Pseudo-3D convolution and attention layers. This design efficiently processes spatiotemporal relationships while optimizing computational resources. The attention mechanism makes use of Pseudo-3D attention layers that apply spatial and temporal attentions in sequence, creating a trade-off between memory efficiency and temporal coherence. Finally, F↑ performs masked frame insertion for free frame rate control and transitions.

\subsubsection{VideoFusion}

\begin{figure}[htbp]
    \centering
    \includegraphics[width=0.65\columnwidth]{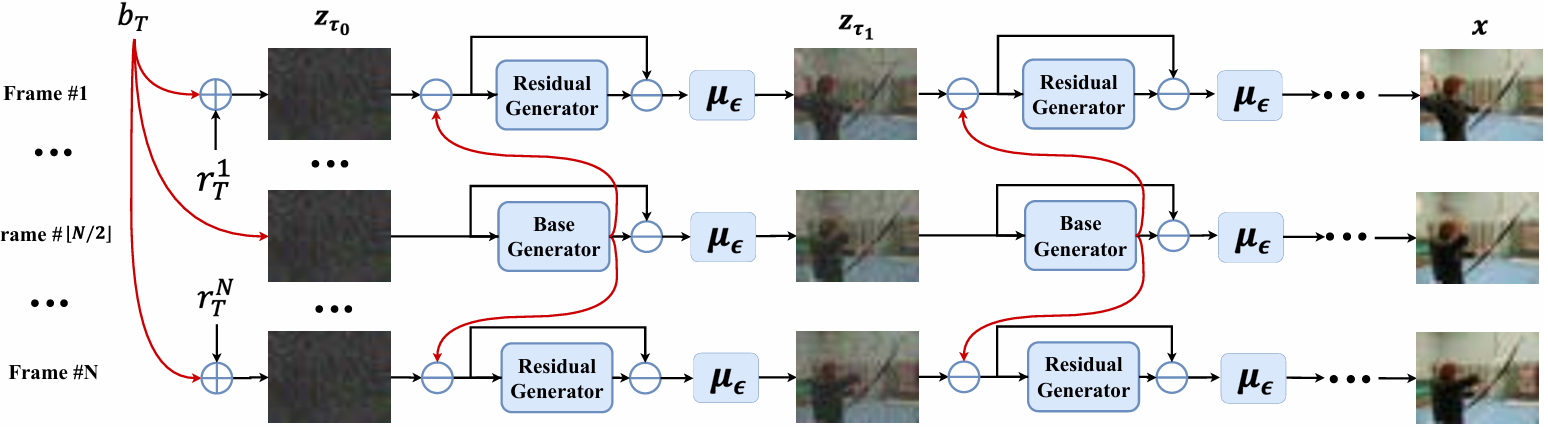}
    \caption{Model Architecture of VideoFusion from~\cite{luo2023videofusion}.}
    \label{fig:videofusion}
\end{figure}

VideoFusion~\cite{luo2023videofusion} proposes a Decomposed Diffusion Probabilistic Model architecture for generating high-quality, temporally coherent video, as shown in Fig.~\figref{fig:videofusion}. It is defined through the decomposition of video synthesis into two complementary generators. It depends on hierarchical noise decomposition, emitting base noise shared between frames and residual noise with only time-specific variations, hence enabling the modeling of both spatial consistency and dynamic frame-to-frame variation. The architecture mainly consists of two parts: (a) a base generator utilizing a pretrained image DPM to handle the static, shared visual content across frames, and (b) a residual generator, introducing temporal details by adding residual noise which adapts the base output to capture the frame-specific motion. During training, both generators are jointly trained to equilibrate the content generated by the pretrained image DPM with temporal refinement from the residual generator so that every frame in the video sequence is coherent with the previous one. The residual generator works complementarily with the base generator to predict a frame-wise noise that selectively changes each frame while keeping the continuous coherence of the base content, balancing computational efficiency with temporal dynamics.

\subsubsection{Latent-Shift}

\begin{figure}[htbp]
    \centering
    \includegraphics[width=0.65\columnwidth]{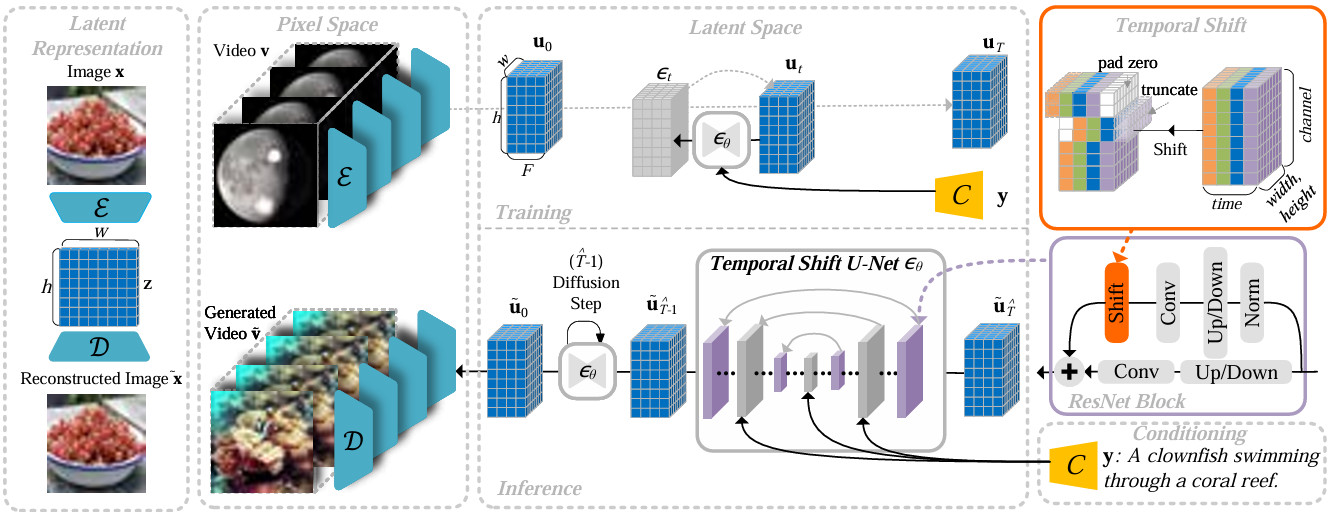}
    \caption{Model Architecture of Latent-Shift from~\cite{an2023latentshift}.}
    \label{fig:latentshift}
\end{figure}

Latent-Shift~\cite{an2023latentshift} adapts a pretrained latent diffusion model~\cite{rombach2022ldm} towards text-to-video generation through a parameter-free temporal shift module included in the model's architecture, as shown in Fig.~\figref{fig:latentshift}. The system first trains a VAE-based autoencoder over images to learn the latent representation and then adapts toward encoding and decoding video frames independently, thus efficiently encoding the spatial information across frames in the latent space. At the heart of Latent-Shift is a Temporal Shift U-Net to learn the denoising of latent video representations. During training, the U-Net processes the latent representation of each frame by sampling different steps of diffusion, and during inference, it progressively denoises from an initial noise distribution until a clean frame sequence is achieved. The architecture is made up of two key modules: 2D ResNet~\cite{he2016resnet} blocks that include convolutional layers, and transformer modules which are additionally modified with spatial attention mechanisms. The proposed temporal shift module is adjusted within the residual branch of each ResNet block, which allows the feature maps to move across the temporal dimension. This setup allows for temporal coherence between frames without requiring additional parameters. Text inputs condition the generated video. With a cross-attention mechanism in transformer blocks, it incorporates text embeddings into spatial attention layers and aligns the generated video content with the provided prompts.

\subsubsection{Free-Bloom}

\begin{figure}[htbp]
    \centering
    \includegraphics[width=0.65\columnwidth]{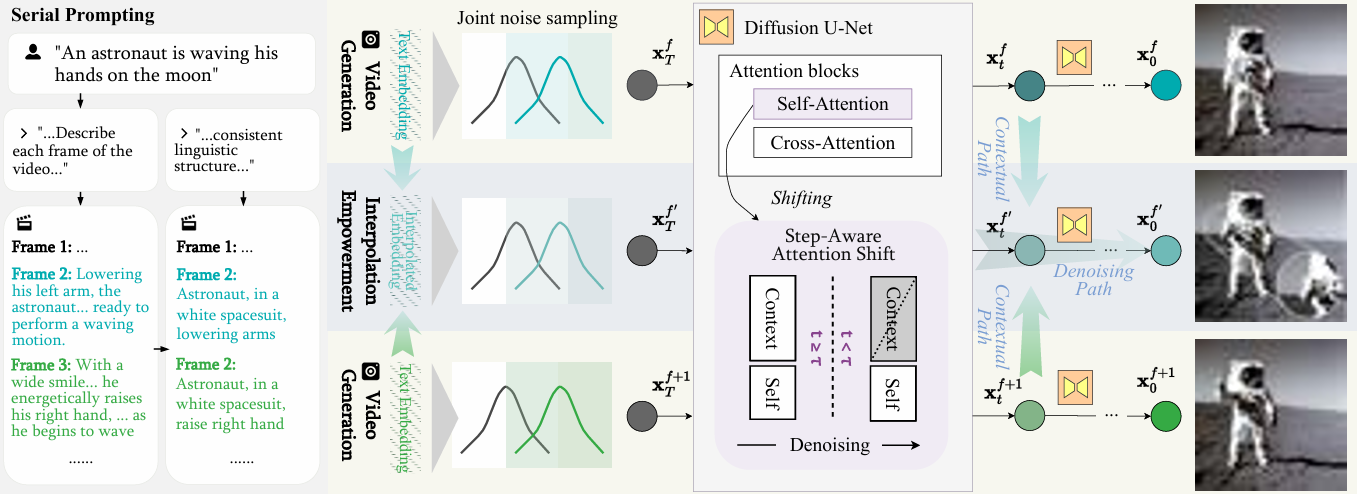}
    \caption{Model Architecture of Free-Bloom from~\cite{huang2023freebloom}.}
    \label{fig:freebloom}
\end{figure}

Free-Bloom~\cite{huang2023freebloom} presents a zero-shot, training-free pipeline with large language models and pretrained latent diffusion models that transform text prompts into high-quality videos without any video data or extra training, as shown in Fig.~\figref{fig:freebloom}. The overall architecture of the model is divided into three sequential modules: serial prompting, video generation, and interpolation empowerment. During the Serial Prompting stage, it is the "director" LLM that turns a single-text prompt into a detailed frame-by-frame serialization describing action development through time while keeping semantic coherence across frames. A pre-trained LDM during Video Generation acts as the "animator": it generates individual frames based on the generated prompt sequence. To achieve the dual goals of temporal and identical coherence, the backward diffusion process is modified in two main ways: first, joint noise sampling, reaching a progressive joint for the noise in order to more effectively balance the content coherence with variation; second, step-aware attention shift, shifting the LDM's U-Net architecture modification of self-attention layers between contextual frames and the current frame, with the variation determined by the denoising time step, to favor identical coherence with semantic alignment. The Interpolation Empowerment module further improves temporal resolution and continuity and interpolates intermediate frames with the Dual-Path Interpolation strategy, comprising a Contextual Path that performs linear interpolation between the latent variables of neighboring frames and a Denoising Path that applies DDIM denoising conditioned on interpolated text embeddings.

\subsubsection{LaVie}

\begin{figure}[htbp]
    \centering
    \includegraphics[width=0.65\columnwidth]{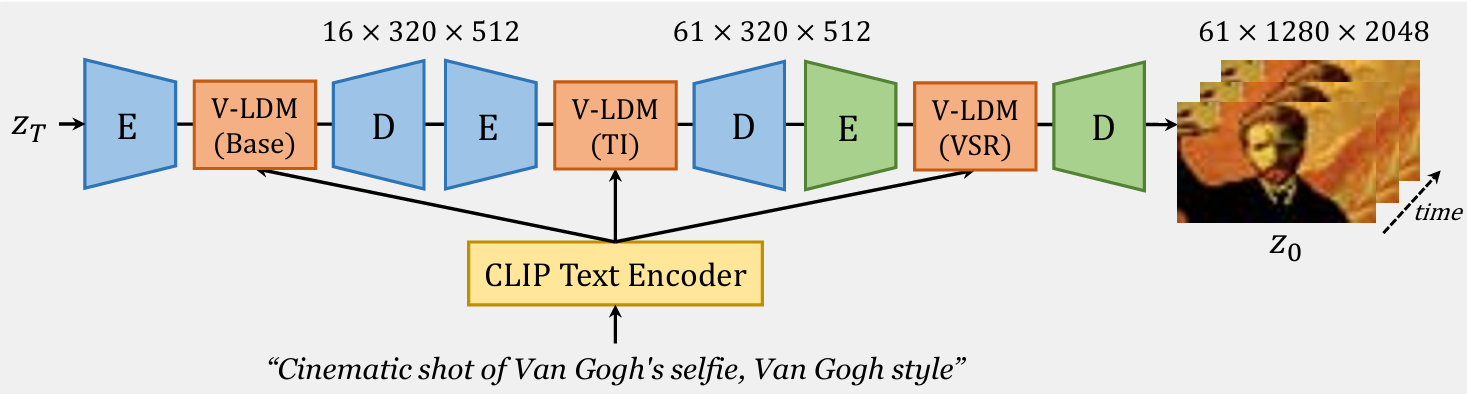}
    \caption{Model Architecture of LaVie from~\cite{wang2023lavie}.}
    \label{fig:lavie}
\end{figure}

LaVie~\cite{wang2023lavie} presents a cascaded framework of Video Latent Diffusion Models (V-LDMs) that leverages pre-trained Text-to-Image (T2I) models for generating high-quality videos, coherent over time, without losing much creative generation capability, as shown in Fig.~\figref{fig:lavie}. This model architecture is ordered into three sequential modules: the Base Text-to-Video (T2V) model, which creates the initial video frames from text input; the Temporal Interpolation (TI) module, which is designed to enhance temporal consistency between frames; and the Video Super-Resolution (VSR) module, which improves the resolution and quality of the generated video. In the Base T2V stage, LaVie extends the pre-trained T2I model, namely Stable Diffusion, to support video data with minimal architectural changes: pseudo-3D convolutions that inflate the 2D kernel to include a temporal dimension and the Spatio-Temporal Transformer (STTransformer) enhanced with Rotary Positional Embedding (RoPE)~\cite{su2021roformer}, capable of capturing rich temporal correlations in video sequences. It jointly trains the model on image and video data to avoid catastrophic forgetting, retaining the richness of creative diversity the T2I model provides. For Temporal Interpolation, the TI model increases temporal resolution by raising the frame rate of the base video. A text-conditioned diffusion U-Net interpolates between keyframes to yield smooth motion with richer temporal detail. In the VSR model, super-resolution of the videos is boosted by incorporating temporal attention along with 3D convolutional layers into a diffusion-based image upscaler. By doing this, the model performs fine-tuning by adjusting the newly added temporal layers while keeping the pre-trained spatial layers fixed—thus improving visual quality while maintaining temporal coherence.

\subsubsection{DreamVideo}

\begin{figure}[htbp]
    \centering
    \includegraphics[width=0.65\columnwidth]{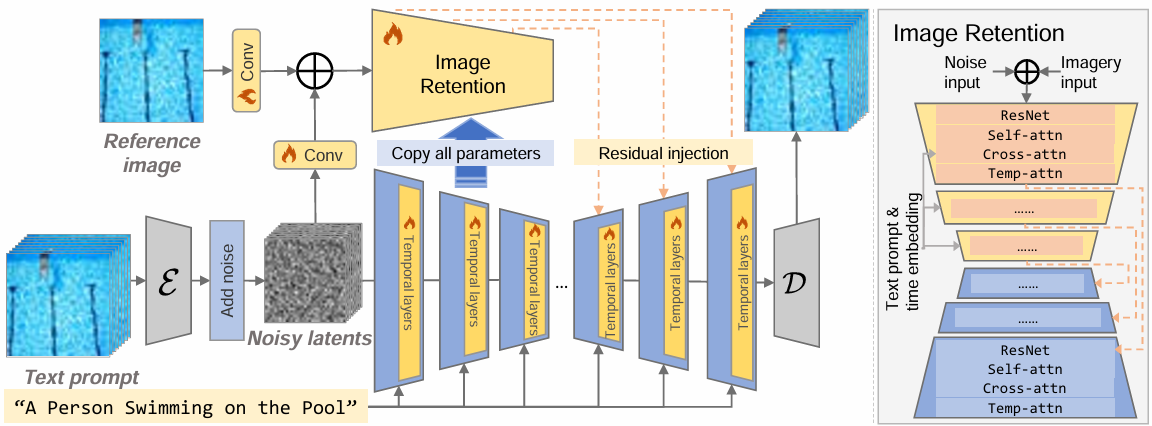}
    \caption{Model Architecture of DreamVideo from~\cite{wei2024dreamvideo}.}
    \label{fig:dreamvideo}
\end{figure}

DreamVideo~\cite{wei2024dreamvideo} presents an image-to-video generation framework that effectively combines an Image Retention block with a pre-trained VLDM for video generation from a single reference image with textual prompts, as shown in Fig.~\figref{fig:dreamvideo}. The main components of the architecture are divided into two parts: the T2V model and the Image Retention block. T2V expands a pre-trained VLDM with Temporal Attention modules (Temp-Attn) and Temporal Convolutional layers (Temp-Conv) inside the U-Net architecture to enable the model to process spatio-temporal relationships in video sequences. Down-sampling and middle layers are followed by the Image Retention block, which adds more convolutional layers to process the input image. It merges the image features into the denoising process by concatenating them with the latent video representations at various levels of U-Net. Classifier-Free Guidance is used with a double session to balance image and text conditions during training and inference. This is done by nullifying each condition independently during training, avoiding overfitting, while enabling adjustable guidance scales during inference to modulate the influence of each modality. The framework also supports two-stage inference to generate longer video sequences or new actions by reusing the final frame of one video as the starting frame for the next with updated prompts.

\subsubsection{Grid-Diffusion}

\begin{figure}[htbp]
    \centering
    \includegraphics[width=0.65\columnwidth]{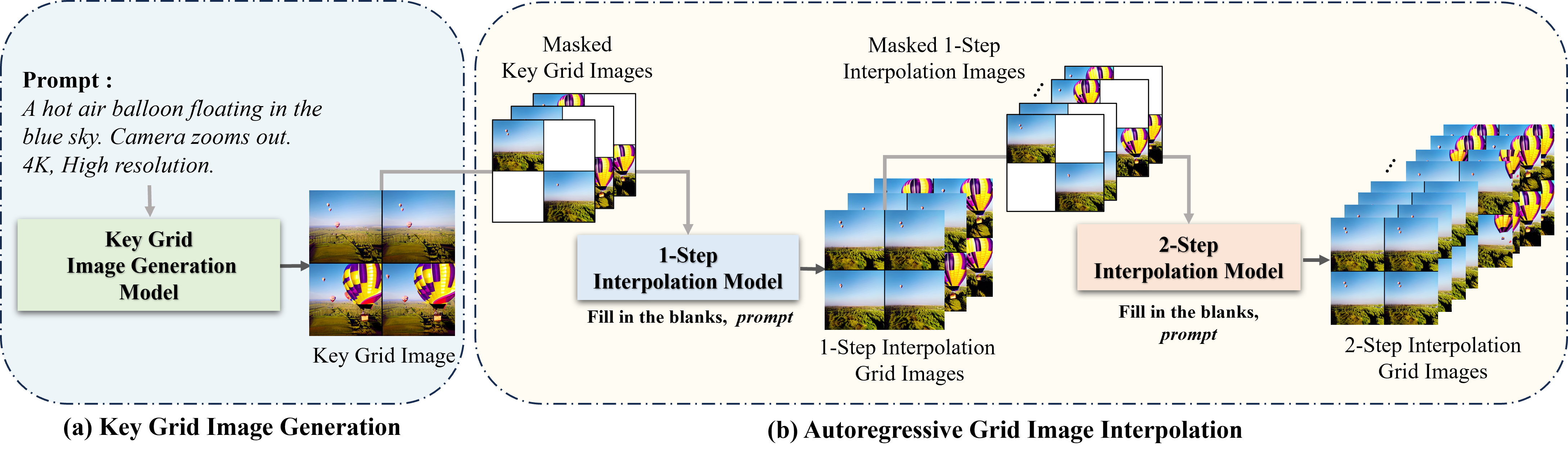}
    \caption{Model Architecture of Grid-Diffusion from~\cite{lee2024grid}.}
    \label{fig:griddiffusion}
\end{figure}

Grid-Diffusion~\cite{lee2024grid} is a diffusion-based framework that reformulates video synthesis as an image generation problem by adopting a grid-based representation, as shown in Fig.~\figref{fig:griddiffusion}. The model is structured around two main modules: keyframe grid generation and autoregressive grid image interpolation. Both modules collaborate to translate textual descriptions into temporally coherent videos. The key grid image generation component uses a fine-tuned Stable Diffusion model to generate a \(2 \times 2\) grid of chronologically ordered frames summarizing the temporal flow of the video. This grid effectively anchors motion and content information using T2I capabilities. The Autoregressive Grid Image Interpolation module contains two sub-models: a 1-Step and a 2-Step Interpolation Model. These operate over masked grid images in an autoregressive fashion, conditioned on previously generated frames, to create intermediate frames and preserve temporal continuity. This setup ensures temporal consistency while avoiding visual artifacts or discontinuities in the output.

\subsubsection{FIFO-Diffusion}

\begin{figure}[htbp]
    \centering
    \includegraphics[width=0.65\columnwidth]{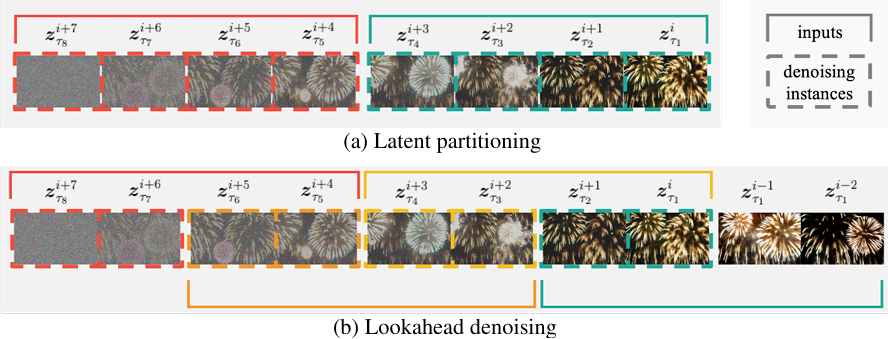}
    \caption{Model Architecture of FIFO-Diffusion from~\cite{kim2024fifodiffusion}.}
    \label{fig:fifodiffusion}
\end{figure}

FIFO-Diffusion~\cite{kim2024fifodiffusion} proposes a training-free inference framework for pre-trained diffusion models that enables infinite-length video generation from a text prompt, as shown in Fig.~\figref{fig:fifodiffusion}. The architecture comprises three main modules: Diagonal Denoising, Latent Partitioning, and Look-ahead Denoising. In the Diagonal Denoising module, frames are processed in a queue-based mechanism following a first-in-first-out (FIFO) order. As fully denoised frames exit the queue, new noisy frames are added, keeping memory usage constant. Latent Partitioning splits the queue into multiple frame blocks to reduce inter-frame noise level variation, improving alignment with the model’s training distribution and allowing parallel computation. Lookahead Denoising enhances forward referencing capabilities by selectively updating frames and incorporating downsampled future frames, giving cleaner context during denoising.

\subsubsection{VideoTetris}

\begin{figure}[htbp]
    \centering
    \includegraphics[width=0.65\columnwidth]{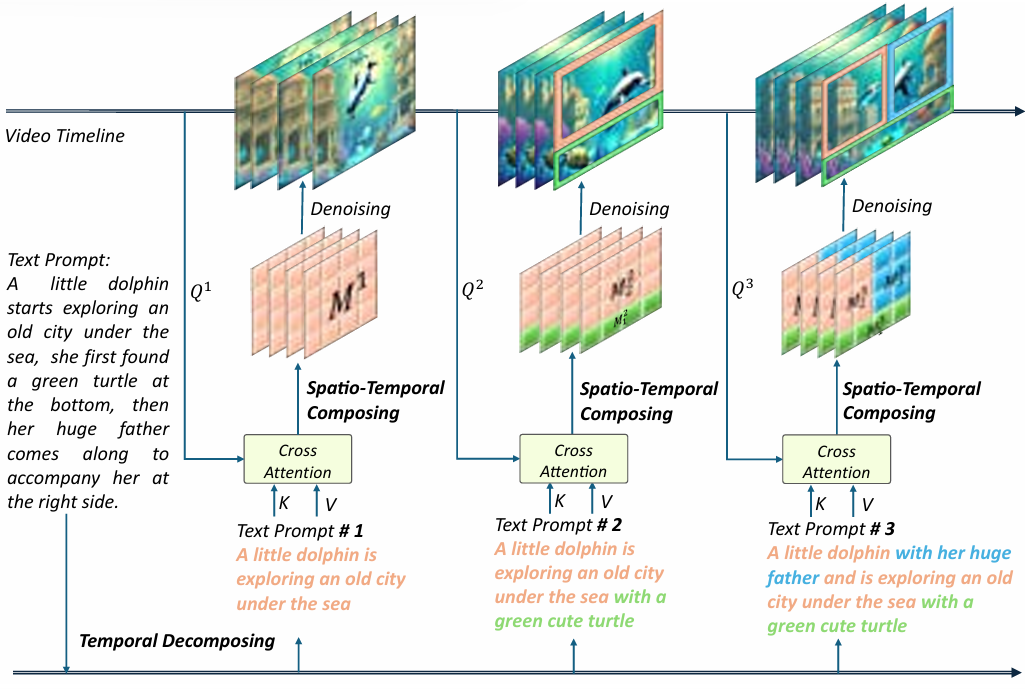}
    \caption{Model Architecture of VideoTetris from~\cite{tian2024videotetris}.}
    \label{fig:videotetris}
\end{figure}

VideoTetris~\cite{tian2024videotetris} is a compositional diffusion-based text-to-video framework capable of generating multi-object videos with control over attributes, spatial layout, and temporal dynamics, as shown in Fig.~\figref{fig:videotetris}. It is composed of three major modules: Spatio-Temporal Compositional Diffusion, Enhanced Video Data Preprocessing, and Consistency Regularization. The Spatio-Temporal Compositional Diffusion module performs: (i) Prompt Decomposition into frame-wise sub-prompts with region masks; (ii) Cross-Attention Manipulation for sub-prompt alignment; (iii) Spatio-Temporal Composition for coherent temporal attention integration. Enhanced Video Preprocessing supports extended video generation through optical flow scoring for motion diversity and language model-based prompt expansion for semantic enrichment. Consistency Regularization ensures object fidelity using reference frame encoding and a reference frame attention mechanism, maintaining spatial integrity across frames via region-specific cross-attention.

\subsubsection{GVDIFF}

\begin{figure}[htbp]
    \centering
    \includegraphics[width=0.65\columnwidth]{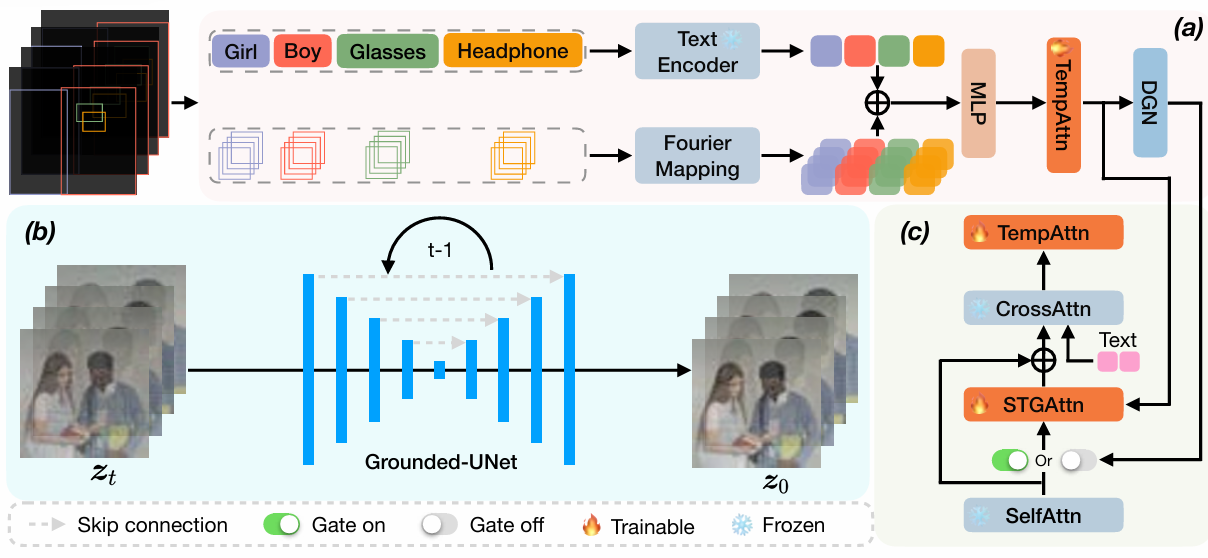}
    \caption{Model Architecture of GVDIFF from~\cite{dou2024gvdiff}.}
    \label{fig:gvdiff}
\end{figure}

GVDIFF~\cite{dou2024gvdiff} is a text-to-video generation framework that integrates both discrete and continuous grounding conditions, as shown in Fig.~\figref{fig:gvdiff}. Its architecture comprises: Uncertainty-Based Grounding Injection, the Spatial-Temporal Grounding Layer (STGL), and a Dynamic Gate Network (DGN), built atop a pre-trained Latent Diffusion Model (LDM). Discrete grounding elements include layouts and keypoints, while continuous ones involve depth and edge maps (e.g., HED, Canny). These are encoded into uncertainty-based representations that guide the U-Net’s attention via modified self-attention layers. The STGL updates transformer layers to handle spatial and temporal grounding, linking grounding features with visual tokens and ensuring alignment across frames. The DGN adaptively skips or engages grounding attention per layer, allowing low-level details in early layers and semantic features in deeper ones. This results in both efficiency and high-fidelity grounded video generation.

\subsubsection{CogVideoX}

\begin{figure}[htbp]
    \centering
    \includegraphics[width=0.5\columnwidth]{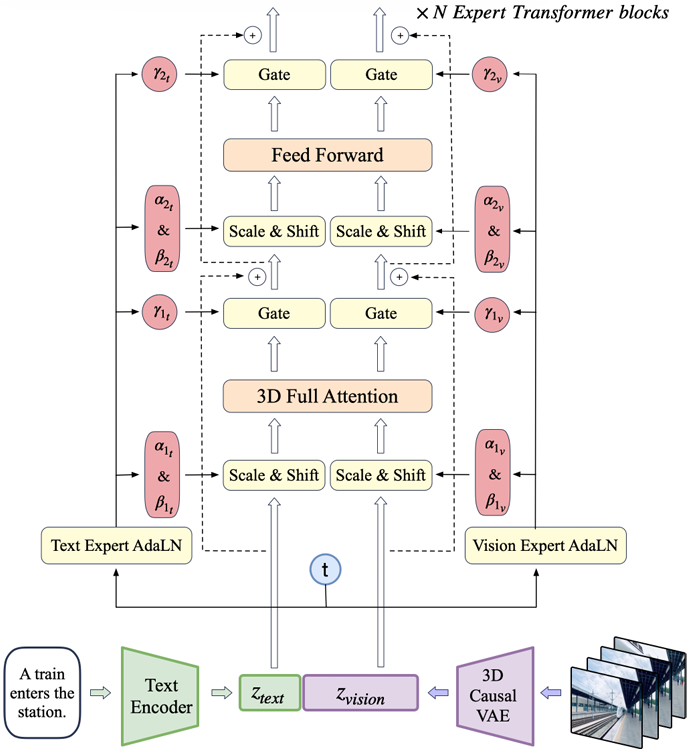}
    \caption{Model Architecture of CogVideoX from~\cite{yang2024cogvideox}.}
    \label{fig:cogvideox}
\end{figure}

CogVideoX~\cite{yang2024cogvideox} implements a diffusion-transformer pipeline for text-to-video generation, converting textual prompts into extended-duration videos without requiring additional video data or training, as shown in Fig.~\figref{fig:cogvideox}. Its architecture comprises several core components: a 3D Causal VAE, an Expert Transformer with Expert Adaptive LayerNorm, and a Progressive Training framework enhanced with Multi-Resolution Frame Packing. The 3D Causal VAE compresses spatial and temporal content through causal convolutions, which ensure that information from future frames is not used in generating past frames—preserving temporal causality. This results in efficient video encoding while maintaining high fidelity. The Expert Transformer fuses modalities by concatenating text embeddings and video latents along the sequence dimension. It employs Expert Adaptive LayerNorm modules that adjust normalization parameters differently for each modality, modulated by diffusion timestep inputs. This aligns the numerical scale and feature space between text and video inputs, enabling effective cross-modal fusion. The Progressive Training framework, combined with Multi-Resolution Frame Packing, enables the model to process videos of varying lengths and resolutions within a single batch. 3D Rotary Positional Embedding (3D-RoPE) is employed to encode relative positions across spatial and temporal dimensions, enhancing coherence and scalability.

\subsubsection{Pyramidal Flow}

\begin{figure}[htbp]
    \centering
    \includegraphics[width=0.65\columnwidth]{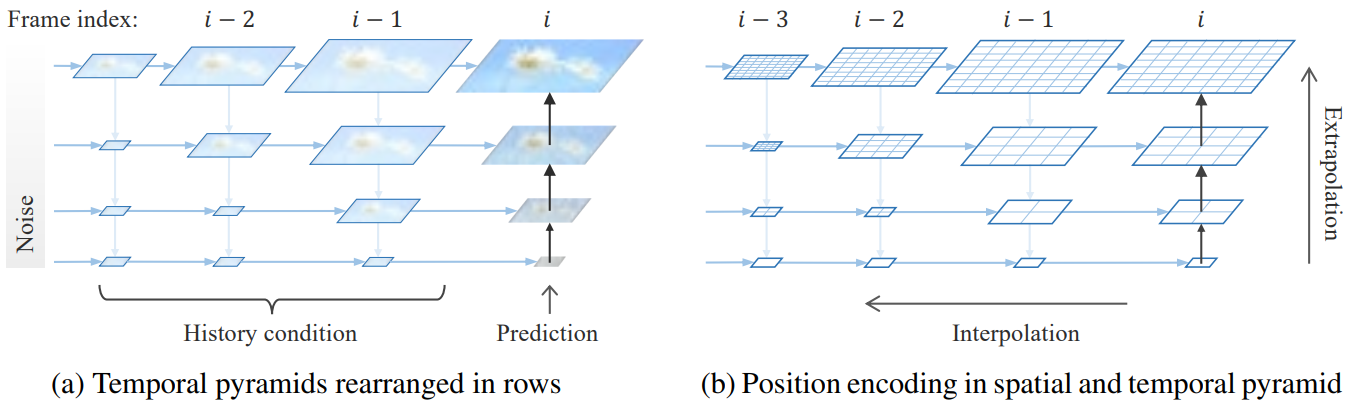}
    \caption{Model Architecture of Pyramidal Flow from~\cite{jin2024pyramidal}.}
    \label{fig:pyramidalflow}
\end{figure}

Pyramidal Flow~\cite{jin2024pyramidal} introduces a diffusion-transformer video generation framework that integrates pyramidal visual representations with flow-matching mechanisms to improve computational efficiency, as shown in Fig.~\figref{fig:pyramidalflow}. The model architecture is built around two main modules: the Spatial Pyramid and the Temporal Pyramid, all within a unified Diffusion Transformer backbone. The Spatial Pyramid decomposes the denoising trajectory spatially across hierarchical stages. Instead of operating on full-resolution video throughout, the model performs denoising at progressively increasing spatial scales, reaching full resolution only in the final pyramid stage. To ensure continuity and consistency across resolutions, Pyramidal Flow Matching is employed to match piecewise flows between noisy and clean latent representations, aligning distributions and enabling smoother transitions between pyramid levels. A Corrective Renoising scheme is applied at each stage to ensure distribution alignment and stability during denoising. The Temporal Pyramid addresses autoregressive video generation's temporal challenges using Pyramidal Temporal Conditioning. It compresses temporal history into progressively coarser representations, which are used as conditioning signals for future frame prediction. This pyramidal temporal structure reduces both memory usage and computational complexity while maintaining temporal fidelity across long sequences.

\section{Training Datasets and Configurations}
\label{sec:datasets}

The diversity and quality of the training datasets, along with proper training configurations, serve as the stepping stones in effectively training text-to-video generation models. The given section provides a summary of the datasets which were utilized to train such T2V models and also provides details of their training configurations.

\subsection{Training Datasets}

The diversity and quality of datasets are of immense importance in developing text-to-video generation models, as it provides foundational data necessary for the training and the performance assessment of such models. Through this section, the paper mentions the datasets used across different papers and gives detailed descriptions of the datasets that aided the training efforts of the various models.

\subsubsection{WebVid-10M}
WebVid-10M~\cite{bain2021frozen} dataset comprises a total of 10.7 million videos tagged with their corresponding text pairs, amounting to close to 52,000 hours of video data. Sourced primarily from stock footage sites such as Shutterstock, each short video is captioned with the respective textual description. First introduced by Bain et al. in the work ``Frozen in Time,'' the dataset allows for joint video and image encoding for retrieval tasks. Due to copyright restrictions, the dataset originally provided URLs and captions for independent downloading; however, as of February 23, 2024, these resources may no longer be available following a cease-and-desist request from Shutterstock.

\subsubsection{UCF-101}
UCF-101~\cite{soomro2012ucf101} dataset comprises close to 13,320 videos which cover actions performed by humans and are classified into 101 classes that range from sports and musical performances to human-object interactions. Extracted from YouTube, it provides a wide range of dynamic camera movements, object appearances, poses, scales, viewpoints, background clutter, and lighting conditions, making it a challenging benchmark for training and evaluating action recognition models.

\subsubsection{HowTo100M}
HowTo100M~\cite{miech2019howto100m} dataset is made up of approximately 136 million video segments created from 1.22 million narrated tutorial videos, covering over 23,000 unique tasks, such as culinary arts, handicrafts, and physical fitness. Each clip includes automatically transcribed subtitles from YouTube, enabling models to learn text-video embeddings from real-world instructional content. Due to copyright restrictions, only URLs and captions were provided for independent downloading, but these resources are no longer accessible as of February 23, 2024.

\subsubsection{LAION-5B}
LAION-5B~\cite{schuhmann2022laion5b} dataset is an extensive collection of 5.85 billion image-text pairs curated to democratize access to large-scale training data for image-text models. It includes 2.3 billion English-language pairs and 2.2 billion pairs in over 100 other languages. Sourced from the Common Crawl repository~\cite{mishne2011commoncrawl}, it focuses on images with associated alt-text descriptions, filtered using OpenAI's CLIP model~\cite{radford2021clip} to retain pairs with high image-text similarity.

\subsubsection{VATEX}
VATEX~\cite{wang2019vatex} dataset is a large-scale multimodal dataset, made up of over 41,250 unique videos. Each video includes annotations with 10 captions in English and 10 in Chinese, resulting in a total of 825,000 captions. The dataset covers a breadth of 600 categories of human activities and supports both multilingual video captioning and video-guided machine translation.

\subsubsection{HD-VILA-100M}
HD-VILA-100M~\cite{xue2022video} dataset includes approximately 100 million high-definition video clips along with their textual descriptions, totaling about 371,500 hours at 720p each. It spans 15 distinct categories, including sports, cooking, and travel, offering rich spatiotemporal contextualization for training video-language models. Each clip averages 13.4 seconds in length with an average of 32.5 words in its description.

\subsubsection{Tai-Chi-HD}
Tai-Chi-HD~\cite{siarohin2019fomm} is a collection of high-resolution videos (180 in total) showing sequences of Tai Chi moves running between 128 and 1,024 frames. The videos, sourced from YouTube, are cropped to focus on complete human bodies performing Tai Chi, serving as a benchmark for high-quality motion transfer and video generation models.

\subsubsection{MUG Facial Expression}
MUG Facial Expression~\cite{aifanti2010mug} dataset contains high-resolution image sequences from 86 subjects performing six basic facial expressions: disgust, fear, anger, happiness, sadness, and surprise. Each sequence includes onset, apex, and offset stages, starting and ending with a neutral expression. The dataset includes 1,462 sequences at 896×896 resolution, along with facial landmark annotations.

\subsubsection{Weizmann Action}
Weizmann Action~\cite{gorelick2005spacetime} dataset consists of 90 low-resolution videos featuring nine subjects performing ten different actions such as bending, jumping, running, and waving. Captured using a static camera and uniform backgrounds, the dataset supports fundamental human action recognition research.

\subsubsection{BAIR}
BAIR~\cite{finn2017video} dataset comprises video sequences of a robotic arm interacting with tabletop objects. It contains about 44,000 sequences of 30 frames each at 64×64 resolution, capturing the dynamics of physical interactions. It has been widely used in action-conditioned video prediction studies.

\subsubsection{TGIF (Tumblr GIF)}
TGIF (Tumblr GIF)~\cite{li2016tgif} dataset is a collection of 100,000 animated GIFs sourced from Tumblr, each captioned with a relevant text description, totaling 120,000 sentences. This dataset supports research in video captioning, retrieval, and text-to-video synthesis by offering animated visual content with associated natural language annotations.

\subsubsection{Moments in Time}
Moments in Time~\cite{monfort2018moments} dataset includes one million videos, each lasting three seconds, labeled with one of 339 action classes involving activities like opening, running, and cooking. Encompassing events involving people, animals, objects, and natural phenomena, it provides both visual and audio information that is highly valuable in the development of models that understand complex events.

\begin{table*}[!t]
\centering
\scriptsize
\renewcommand{\arraystretch}{1.2}
\setlength{\tabcolsep}{4pt}
\caption{Training Configurations of Survey Models}
\label{tab:training_config_compact}
\resizebox{\textwidth}{!}{%
\begin{tabular}{c|c|c|c|c|c|c|c|c}
\hline
\textbf{Model Name} & \textbf{GPU} & \textbf{GPU Count} & \textbf{Batch Size} & \textbf{Learning Rate} & \textbf{Optimizer} & \textbf{Steps/Epochs} & \textbf{Loss Function} & \textbf{Dropout Rate} \\
\hline
MoCoGAN~\cite{tulyakov2018mocogan} & - & - & -   & -               & Adam  & -                       & Adversarial           & -    \\
VideoGPT~\cite{yan2021videogpt}    & - & - & 128 & $1\times10^{-4}$ & Adam  & -                       & Cross-entropy         & -    \\
GODIVA~\cite{wu2021godiva}         & V100 & 64 & 512 & $1\times10^{-4}$ & Adam  & 200k steps              & Cross-entropy         & -    \\
N\"UWA~\cite{wu2021nuwa}             & - & - & -   & -               & -     & -                       & -                     & -    \\
CogVideo~\cite{hong2022cogvideo}   & A100 & 32 & 256 & $1\times10^{-4}$ & Adam  & 1M steps                & Cross-entropy         & -    \\
Make-A-Video~\cite{singer2022makeavideo} & A100 & 2048 & - & -          & -     & -                       & Standard diffusion    & -    \\
VideoFusion~\cite{luo2023videofusion} & - & - & - & -               & -     & -                       & -                     & -    \\
Latent-Shift~\cite{an2023latentshift} & - & - & 256 & $1\times10^{-5}$ & Adam  & -                       & MSE (latent space)    & 0    \\
Free-Bloom~\cite{huang2023freebloom} & RTX 3090 Ti & 1 & - & -         & -     & -                       & -                     & -    \\
LaVie~\cite{wang2023lavie}         & - & - & -   & -               & -     & -                       & MSE (latent space)    & -    \\
DreamVideo~\cite{wei2024dreamvideo} & - & 8 & 9   & -               & -     & 5 epochs (340k samples) & MSE                   & -    \\
Grid Diffusion~\cite{lee2024grid}  & A100 80GB & 2 & 28/20 & -        & -     & 82k/54k steps           & -                     & -    \\
FIFO-Diffusion~\cite{kim2024fifodiffusion} & - & - & - & -          & -     & -                       & -                     & -    \\
VideoTetris~\cite{tian2024videotetris} & A800 & 4 & 2 & $1\times10^{-5}$ & -  & 16k steps               & -                     & 0.05 \\
GVDIFF~\cite{dou2024gvdiff}        & - & - & -   & -               & -     & -                       & -                     & -    \\
CogVideoX~\cite{yang2024cogvideox} & - & - & -   & -               & -     & -                       & L1+LPIPS+KL+GAN       & -    \\
Pyramidal Flow~\cite{jin2024pyramidal} & A100 & 128 & 1536/768/384 &
$1\times10^{-4},\; 5\times10^{-5}$ & AdamW & 300k total steps & Flow matching & - \\
\hline
\end{tabular}%
}
\end{table*}

\subsection{Training Configurations}

Understanding the complexities of training configurations becomes essential for researchers entering the field of text-to-video generation. These configurations serve as practical guidelines for evaluating the feasibility of implementing or enhancing existing approaches under specific resource constraints. Familiarity with computational requirements, optimization strategies, and architectural decisions not only aids reproducibility but also highlights potential bottlenecks and opportunities for improvement. Table~\tabref{tab:training_config_compact} catalogs the essential training parameters reported across surveyed models, including GPU type and count, batch sizes, learning rates, optimizer choices, training steps or epochs, loss functions, and dropout rates. Together, these parameters offer a comprehensive overview of the computational and architectural needs required to train text-to-video models. They are thus crucial reference points for future research and implementation planning. While several papers do not provide complete training configurations~\cite{wu2021nuwa, hong2022cogvideo, singer2022makeavideo, luo2023videofusion, an2023latentshift, huang2023freebloom, kim2024fifodiffusion}, insights can still be extracted through available metadata, ablation studies, and cross-referencing within the papers. These partial details, along with identifiable patterns across model families, allow for preliminary estimates of resource needs and can help guide researchers in assessing implementation feasibility under various computational budgets.

\section{Performance Evaluation}
\label{sec:evaluation}

The fast development of text-to-video generation models calls for an equally strong framework that provides evidence for their effectiveness and applicability in real-world applications. The performance evaluation includes both quantitative metrics for objective measurement over various generated video aspects and qualitative assessments, often through human evaluations, for capturing perceptual qualities not satisfactorily addressed by metrics alone. Added to that, new benchmarks are being developed that tackle the changing challenges in the field, which allow for more comprehensive and varied evaluations. The current section will give an overview of the current standard evaluation metrics and benchmarks, discuss the place of human evaluations in qualitative assessment, and review one of the newer benchmarks that seek to provide more nuanced and comprehensive evaluations of model performance.

\subsection{Current Standard Evaluation Metrics and Benchmarks}

\subsubsection{Evaluation Metrics}
Evaluation of generative video models depends on a set of currently standardized, quantitative metrics that objectively assess generated content. Such uniformity in evaluation metrics facilitates benchmarking and monitoring progress in respect of such generative models. The most widely adopted quantitative metrics are:

\textbf{(i) Inception Score \textit{(IS)}}~\cite{salimans2016improved}: The Inception Score provides quality and diversity assessments of generated videos by employing a pre-trained Inception network~\cite{szegedy2015inception} used to classify frames. The higher the \textit{IS} is, the better the quality and diversity of videos are, hence making the generated easily recognizable by the classification model, denoted in Eq.~(\Eqref{eq:is}):
\begin{equation}
\mathit{IS} = \exp\left(\mathbb{E}_x \left[ D_{\mathit{KL}} \left( p(y|x) \, \| \, p(y) \right) \right] \right)
\label{eq:is}
\end{equation}

Here, $p(y|x)$ is the probability distribution over classes $\mathit{y}$ for a generated video $\mathit{x}$, and $p(y)$ is the marginal distribution over classes. $D_{\mathit{KL}}$ denotes the Kullback-Leibler divergence~\cite{kullback1951information}, measuring the difference between these distributions.

\textbf{(ii) Fréchet Inception Distance \textit{(FID)}}~\cite{heusel2017ttur}: \textit{FID} measures the similarity between the distributions of generated and real videos by comparing their feature representations. Originally designed for images, \textit{FID} captures both quality and diversity, serving as a foundation for the Fréchet Video Distance (\textit{FVD}), denoted in Eq.~(\Eqref{eq:fid}):

\begin{equation}
\mathit{FID} = \left\| \mu_g - \mu_r \right\|^2 + \mathrm{Tr} \left( \Sigma_g + \Sigma_r - 2 \left( \Sigma_g \Sigma_r \right)^{1/2} \right)
\label{eq:fid}
\end{equation}

In this formula, $\mu_g$ and $\mu_r$ are the mean feature vectors of generated and real videos, respectively. $\Sigma_g$ and $\Sigma_r$ are their covariance matrices. $\mathrm{Tr}$ denotes the trace of a matrix, representing the sum of its diagonal elements.

\textbf{(iii) Fréchet Video Distance \textit{(FVD)}}~\cite{unterthiner2019fvd}: \textit{FVD} assesses both quality and temporal coherence by comparing feature distributions over time. A lower \textit{FVD} score indicates that generated videos closely resemble real videos in both visual quality and motion dynamics, denoted in Eq.~(\Eqref{eq:fvd}):
\begin{equation}
\mathit{FVD} = \left\| \mathit{\mu}_g - \mathit{\mu}_r \right\|^2 + \mathrm{Tr} \left( \mathit{\Sigma}_g + \mathit{\Sigma}_r - 2 \left( \mathit{\Sigma}_g \mathit{\Sigma}_r \right)^{1/2} \right)
\label{eq:fvd}
\end{equation}

Here, $\mathit{\mu}_g$ and $\mathit{\mu}_r$ represent the mean feature vectors, while $\mathit{\Sigma}_g$ and $\mathit{\Sigma}_r$ are the covariance matrices of the generated and real videos. The term $\left( \mathit{\Sigma}_g \mathit{\Sigma}_r \right)^{1/2}$ is the matrix square root of the product of the covariance matrices.

\textbf{(iv) CLIP Similarity \textit{(CLIP-SIM)}}~\cite{radford2021clip}: \textit{CLIP-SIM} measures how well the generated videos align semantically to their provided corresponding text descriptions, utilizing the CLIP model~\cite{radford2021clip} for this purpose. A higher \textit{CLIP-SIM} score indicates better semantic coherence between video content and text prompts, denoted in Eq.~(\Eqref{eq:clip_oa}):
\begin{equation}
\mathit{CLIP\text{-}SIM}
= \frac{\left\langle \mathrm{CLIP}(V),\, \mathrm{CLIP}(T) \right\rangle}
{\left\| \mathrm{CLIP}(V) \right\| \, \left\| \mathrm{CLIP}(T) \right\|}
\label{eq:clip_oa}
\end{equation}

In this equation, $\mathrm{CLIP}(V, T)$ is the cosine similarity between video $\mathit{V}$ and text prompt $\mathit{T}$ embeddings from the CLIP model. $\left\| \mathrm{CLIP}(V) \right\|$ and $\left\| \mathrm{CLIP}(T) \right\|$ are the Euclidean norms of the video and text embeddings, respectively.

\textbf{(v) Kernel Video Distance \textit{(KVD)}}~\cite{unterthiner2018video}: \textit{KVD} evaluates the similarity between the feature distributions of real and generated videos, focusing on both spatial and temporal consistency. It is based on the Maximum Mean Discrepancy (\textit{MMD}), applied to features extracted from pre-trained networks that capture spatiotemporal information in video data. A reduced \textit{KVD} score means that the created video aligns well with real video data, both in terms of visual appearance and motion fluidity. First, $\mathit{MMD}^2$ is calculated as, denoted in Eq.~(\Eqref{eq:mmd}):

\begin{equation}
\begin{aligned}
\mathit{MMD}^2(\mathit{X}, \mathit{Y}) &= \frac{1}{\mathit{m}(\mathit{m}-1)} \sum_{i \neq j} k(\mathit{x}_i, \mathit{x}_j) \\
&\quad + \frac{1}{\mathit{n}(\mathit{n}-1)} \sum_{i \neq j} k(\mathit{y}_i, \mathit{y}_j) \\
&\quad - \frac{2}{\mathit{m}\mathit{n}} \sum_{i=1}^{\mathit{m}} \sum_{j=1}^{\mathit{n}} k(\mathit{x}_i, \mathit{y}_j)
\end{aligned}
\label{eq:mmd}
\end{equation}

In this equation, $\mathit{X} = \{\mathit{x}_1, \mathit{x}_2, \dots, \mathit{x}_m\}$ represents feature vectors gathered from real videos, while $\mathit{Y} = \{\mathit{y}_1, \mathit{y}_2, \dots, \mathit{y}_n\}$ corresponds to feature vectors from generated videos. The kernel function $k(\mathit{x}, \mathit{y})$ (e.g., Gaussian) measures similarity between feature vectors. Finally, \textit{KVD} is obtained by taking the square root of $\mathit{MMD}^2$, denoted in Eq.~(\Eqref{eq:kvd}):

\begin{equation}
\mathit{KVD} = \sqrt{\mathit{MMD}^2(\mathit{X}, \mathit{Y})}
\label{eq:kvd}
\end{equation}

A reduced \textit{KVD} score ensures higher quality in the created video, translating to a realistic visual appearance and smooth motion alignment with actual video data.

Performance evaluation of these models on the mentioned metric~\cite{salimans2016improved, unterthiner2019fvd, hossain2024simclip} and datasets is provided in Table~\tabref{tab:eval_metrics}.

\begin{table*}[!t]
\centering
\caption{Performance Comparison of Text-to-Video Generation Models on Standard Benchmarks}
\label{tab:eval_metrics}
\renewcommand{\arraystretch}{1.3}
\setlength{\tabcolsep}{4pt}
\resizebox{\textwidth}{!}{%
\begin{tabular}{c|c|c|c|c|c|c|c|c|c}
\hline
\multicolumn{1}{c|}{\multirow{2}{*}{\textbf{Model Name}}} &
\multicolumn{3}{c|}{\textbf{UCF-101}} &
\multicolumn{3}{c|}{\textbf{MSRVTT}} &
\textbf{BAIR} &
\multicolumn{2}{c}{\textbf{Kinetics-600}} \\
\cline{2-10}
& \textbf{IS} ($\uparrow$) & \textbf{FVD} ($\downarrow$) & \textbf{CLIPSIM} ($\uparrow$)
& \textbf{IS} ($\uparrow$) & \textbf{FVD} ($\downarrow$) & \textbf{CLIPSIM} ($\uparrow$)
& \textbf{FVD} ($\downarrow$) & \textbf{FVD} ($\downarrow$) & \textbf{CLIPSIM} ($\uparrow$) \\
\hline
MoCoGAN~\cite{tulyakov2018mocogan}         & 12.48$\pm$0.03 & - & - & - & - & - & 503   & - & - \\
VideoGPT~\cite{yan2021videogpt}           & 24.69$\pm$0.30 & - & - & - & - & - & 103.3 & - & - \\
GODIVA~\cite{wu2021godiva}                & -              & - & - & - & - & 0.2402 & -    & - & - \\
N\"UWA~\cite{wu2021nuwa}                    & -              & - & - & - & - & 0.2439 & -    & - & - \\
CogVideo~\cite{hong2022cogvideo}          & 50.46          & 626 & 0.3025 & 1294 & - & 0.2631 & - & - & - \\
Make-A-Video~\cite{singer2022makeavideo}  & 33             & 367.23 & - & 550 & - & 0.3049 & 86.9 & - & 0.3012 \\
VideoFusion~\cite{luo2023videofusion}     & 71.67          & 139 & - & - & - & 0.293 & 109.23 & - & - \\
Latent-Shift~\cite{an2023latentshift}     & -              & - & - & - & - & 0.2773 & - & - & - \\
Free-Bloom~\cite{huang2023freebloom}      & -              & - & - & - & - & - & - & - & - \\
LaVie~\cite{wang2023lavie}                & -              & - & - & - & - & 0.2949 & - & - & - \\
DreamVideo~\cite{wei2024dreamvideo}       & 54.39          & 197.66 & - & 15.25 & 149.18 & - & - & - & - \\
Grid Diffusion~\cite{lee2024grid}         & 62.88          & 340 & 0.3282 & 375 & - & 0.3096 & - & - & - \\
FIFO-Diffusion~\cite{kim2024fifodiffusion}& -              & - & - & - & - & - & - & - & - \\
VideoTetris~\cite{tian2024videotetris}    & -              & - & - & - & - & - & - & - & - \\
GVDIFF~\cite{dou2024gvdiff}               & -              & - & - & - & - & - & - & - & - \\
CogVideoX~\cite{yang2024cogvideox}        & -              & - & - & - & - & - & - & - & - \\
Pyramidal Flow~\cite{jin2024pyramidal}    & -              & - & - & - & - & - & - & - & - \\
\hline
\end{tabular}%
}
\end{table*}

\subsubsection{Evaluation Benchmarks}
The systematic evaluation of text-to-video generation models needs some standardized datasets that can very effectively assess the technical capabilities and the creative aspects of video synthesis. These benchmark datasets are crucial testing grounds, enabling quantitative comparison across different architectures while challenging models across diverse scenarios, temporal dynamics, and semantic complexity. The following benchmarks are the current standard evaluation frameworks, each of which targets specific aspects of video generation competency, from action recognition to temporal coherence and semantic alignment.

\textbf{(i) UCF-101}~\cite{soomro2012ucf101}: UCF-101 contains 13,320 video clips covering a total of 101 action categories. These action categories include sports, daily actions, and other complex motions, challenging models to generate coherent and contextually relevant videos across various action types.

\textbf{(ii) MSR-VTT}~\cite{xu2016msrvtt}: The dataset is made up of close to 10,000 videos which are tagged with their corresponding 200,000 textual descriptions. It is made in a way to evaluate how well the models can align generated video content with provided textual prompts, allowing for the assessment of semantic coherence and content relevancy.

\textbf{(iii) Kinetics-600}~\cite{carreira2018kinetics}: Kinetics-600 contains over 500,000 video clips across 600 action classes. This thorough dataset covers a broad range of activities, testing models' ability to generate detailed and dynamic videos that accurately show a wide variety of complex actions.

\textbf{(iv) BAIR}~\cite{finn2017video}: The dataset comprises of video sequences of a robotic arm interacting with objects on a tabletop. Its training data consists of approximately 44,000 video sequences, each with 30 frames in resolution 64×64 pixels that capture the dynamic relations between robots and objects. It has been used in video prediction and action-conditioned video generation research, furnishing the models with a baseline to learn the dynamics of physical interactions.

\subsection{Human Evaluations}

While quantitative metrics provide measures objectively, they just cannot capture subjective quality aspects of the generated video, which are crucial for user satisfaction and real-world usage. Thus, human evaluations, in different forms and metrics, have been introduced in multiple papers to assess characteristics like visual realism, semantic consistency, and general aesthetic appeal. They hint at how well a model meets the expectations or preferences of a human being. Human evaluations are purely subjective methods of qualitatively assessing the dimensions of the generation models of videos that capture those aspects, such as realism, coherence, semantic alignment, and others, which could be missed by automated metrics. These ensure that the generated videos meet human expectations and preferences, thus confirming that models produce content that is both technically proficient and subjectively satisfying. This section outlines the four main core human evaluation metrics that have been used across various papers to measure video quality based on human rankings: Fidelity to Text/Alignment to Text, Motion Realism, Aesthetic/Quality, and Overall Quality/Preference.

\subsubsection{Fidelity to Text/Text Alignment}
Fidelity to Text, also at times referred to as Alignment to Text, refers to the degree of representation that the generated video exhibits with respect to the input textual description. This metric ensures not only that the content, context, and actions shown in the video are semantically consistent but also that they are appropriate to the given prompts. Different papers use different terminologies to present this metric. For example, some papers refer to it as ``Semantic Relevance''~\cite{yan2021videogpt}, whilst others refer to this metric as ``Semantic Consistency (SC)''~\cite{wu2021godiva}, ``Faithfulness''~\cite{singer2022makeavideo}, ``Instruction Following''~\cite{yang2024cogvideox}, and ``Fidelity to Text''~\cite{dou2024gvdiff}.

\subsubsection{Realism of Motion}
Motion Realism measures the coherence and smoothness of the motion in the generated video, which should be natural, logical, and continuous in both action and transition, with a fast change or transition of frames. This metric is important to assess the believability and engagement of the motion of objects and subjects within the videos. Papers introduce this metric with various terms and different measures. Metrics ``Motion Realism''~\cite{yan2021videogpt}, ``Plausibility of Motion''~\cite{luo2023videofusion, dou2024gvdiff}, ``Temporal Coherence''~\cite{huang2023freebloom}, ``Motion Awareness Evaluation (MAWE)''~\cite{tian2024videotetris}, and ``Motion''~\cite{lee2024grid, kim2024fifodiffusion} are some of the other names by which this metric goes by across various papers.

\subsubsection{Aesthetic/Quality}
Aesthetic Quality or Quality evaluations are subjective measures of the generated video's visual appeal, including the richness of senses involved and overall stylistic coherence. It includes information about colour harmony, the quality of textures appearing in the video, and artistic expressions. Other names by which this metric goes in various papers are: ``Visual Realisticity''~\cite{wu2021godiva}, ``Aesthetic Quality''~\cite{luo2023videofusion, tian2024videotetris, yang2024cogvideox, lee2024grid, kim2024fifodiffusion, wang2023lavie}, ``Sensory Quality'' and ``Cover Quality''~\cite{yang2024cogvideox}.

\subsubsection{Overall Quality/Preference}
Overall Quality/Preference is used to serve as a thorough measure for the generated video to assess the overall desirability and visual excellence from a human perspective. This metric typically encompasses various qualitative dimensions to determine the general desiredness of the videos. Different Papers refer to this metric using different terminologies. For instance, some of the papers refer to it as ``User Preference Score''~\cite{tulyakov2018mocogan}, whereas other papers use the terms like ``Overall Quality Score''~\cite{yan2021videogpt}, ``Visual Realisticity''~\cite{wu2021godiva}, ``Overall Rank''~\cite{huang2023freebloom}, and ``Overall Preference''~\cite{singer2022makeavideo, luo2023videofusion, tian2024videotetris, dou2024gvdiff, yang2024cogvideox, wei2024dreamvideo, kim2024fifodiffusion}.

\subsection{Towards A Standardized Video Generation Evaluation Benchmark}

\begin{figure}[htbp]
    \centering
    \includegraphics[width=\linewidth]{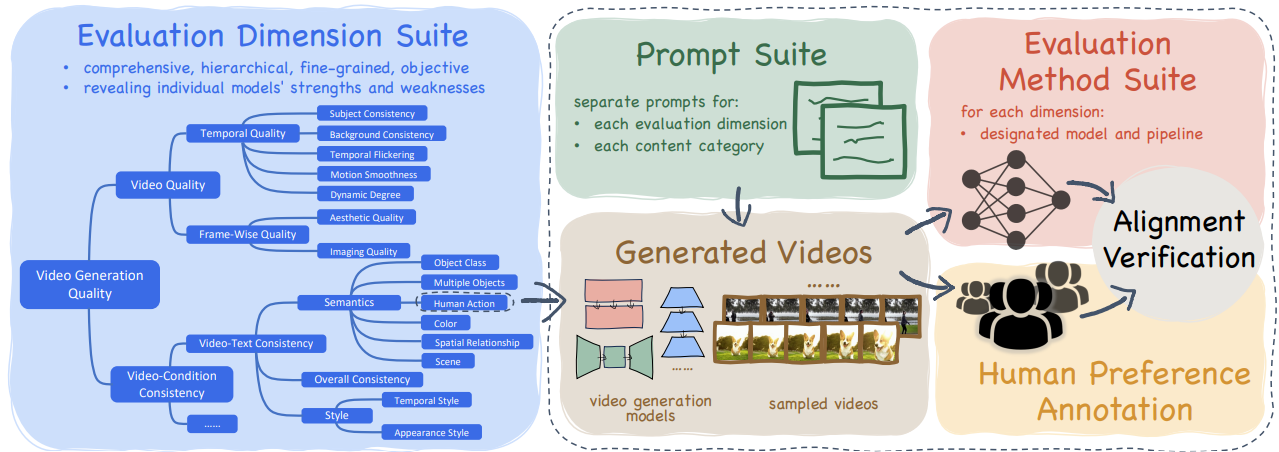}
    \caption{Evaluation Pipeline of VBench from~\cite{huang2024vbench}}
    \label{fig:vbench}
\end{figure}

While being widely adapted, the conventional metrics like Inception Score~\cite{salimans2016improved}, Fréchet Inception Distance~\cite{heusel2017ttur}, Fréchet Video Distance~\cite{unterthiner2019fvd}, and CLIPSIM~\cite{radford2021clip} are limited in the holistic assessment of video generative models. They mostly focus on the statistical similarities between generated and real data distributions, while they fundamentally cannot capture key qualitative aspects that match human perception. Such a simplification of video quality to single numerical scores not only avoids the rich complexity of video assessment but also does not provide sufficient granularity to allow for effective model evaluation. Metrics initially designed for static images and traditional video analysis cannot capture challenges particular to video, such as temporal coherence or semantic consistency. More critically, they are not able to quantify key human-centered characteristics such as identity preservation, motion fluidity, and temporal stability, which are essential for any standard quality assessment in video. To address these limitations, VBench introduces an extensive benchmarking methodology for generative video models~\cite{huang2024vbench}. This framework systematically decomposes video quality assessment into 16 different dimensions of evaluation, offering a fine-grained assessment of video generation quality closer to human perception. By surpassing traditional single-score metrics, VBench supplies researchers with diagnostic tools and, with the inclusion of human preference annotations, allows for the identification of strengths and weaknesses of a model, fostering focused improvements in video generation capabilities, detailed in Fig.~\figref{fig:vbench}.

The assessment process of this framework is based on its hierarchically organized Evaluation Dimension Suite, which classifies the various assessment criteria into two large domains: Video Quality and Video-Condition Consistency. For Video Quality, this framework evaluates several important temporal dimensions, including subject consistency, background stability, temporal flickering, motion smoothness, and dynamic degree, along with frame-like attributes that involve aesthetic and imaging qualities. At the semantic level, Video-Condition Consistency evaluations include object accuracy, multi-object composition, correctness of human actions, consistency in colors, relationships among objects in space, scene correctness, adherence to style, and alignment between video and its corresponding text. VBench follows a comprehensive Prompt Suite of about 100 creatively designed prompts across eight different content categories to ensure that the evaluation is thorough. The framework uses specialized evaluation methods for background consistency, and motion priors for smoothness evaluation, to objectively measure each quality aspect. These automated evaluations are validated through systematic human preference annotations: annotators compare video pairs across specific dimensions, ensuring alignment with human perception. Given the holistic and systematic nature of the review, VBench has the potential to become the go-to standard framework for evaluation of T2V models. Additionally, because of its thorough human validation, its detailed assessment method can show a stronger correspondence with viewers' perceptions, making it valuable for real-world applications. VBench allows for fair model comparisons across the research community and promotes consistent practices for model evaluations through open sourcing and extensibility. Its multi-dimensional approach serves as both an analytical tool and a guide for future research, allowing developers to focus on selected quality aspects most relevant to their use case. This versatility, coupled with its comprehensive methodology, positions VBench as a foundational framework that could guide the development and standardization of future video generation evaluation technology.

\section{Future Direction}
\label{sec:future}

While Text-to-Video Generation has demonstrated some encouraging progress through deep learning approaches, most of the fundamental challenges remain open. The existing models, reporting initial success in generating simple video clips from text descriptions, still struggle with their computational efficiency, dataset limitations, and output quality. In this section, the article discusses some of the possible research avenues that may help overcome these challenges. It also considers the evolution of the creation process of datasets, model architectures, and application domains. Although the suggestions are far from exhaustive, they do hope to contribute something to the oncoming discussion about the progression of such text-to-video technology.

\subsection{Dataset Enrichment}

Currently available datasets for T2V synthesis have a severe weakness because of their limited size, quality, or legal conditions restricting the development of strong models capable of generalization outside a particular setting to generate high-quality video. One possible avenue of research might be the synthesis of datasets using game engines like Unity or Unreal Engine in order to create large-scale, high-resolution, and diverse datasets with no copyright infringement. It does this by developing a generalized prompt framework for overall development, classifying textual descriptions into principal subject, adjectives and attributes, verbs and actions, background, theme and style, and metadata of the camera. Generalization hierarchically allows comprehensive coverage of the subject matter, like people, animals, and objects while avoiding biases due to the use of different attributes and settings. These can range from “a person running through a futuristic cityscape with cyberpunk aesthetics” to “an animal exploring a fantasy forest in anime style.” Feeding such structured prompts into game engines allows the generation of videos corresponding to each scenario with broad avenues for customization. Automating the pipelines of these game engines will efficiently generate large-scale datasets, improving quality and quantity. This approach would not only overcome any kind of limitation that might have existed in existing datasets but also keep it ethical and legal, avoiding any use of copyrighted material. Dataset augmentation this way may finally allow the training of models that generate semantically correct and visually realistic videos— which is a major push forward within the realm of text-to-video synthesis.

\subsection{Model Architectures and Optimization}

Text-to-video generation models currently face several significant limitations that hinder their effectiveness and scalability. Training is computationally intensive because the current architectures are not optimized to sequentially process video data efficiently. Novel model architectures or algorithms that can better handle temporal sequences therefore need to be investigated to achieve more economy in resource usage and training times. These models also tend to produce only short output videos, which is highly limiting for scenarios with longer content. Hence, in the future, research is required for allowing the generation of longer videos by enhancing these models’ temporal modelling. It is also limited to the generation of content with little diversity, partly because the training data itself lacks variety, making the models unable to create wide variations of scenes or subjects. This could be improved with the addition of more diverse datasets or the use of advanced augmentation techniques. Moreover, most generated videos suffer from temporal and spatial incoherence: objects often shift, appear, or disappear unnaturally. Such unrealistic simulations make viewers break immersion; hence, there is a need for further development of temporal consistency and spatial coherence by better modelling of the sequence and imposing physical constraints. The interaction of physics also, in most cases, is unrealistic, hence making such generated videos less plausible; there is great a need for models to understand and, consequently do a better job of simulating the physics of real-world interactions. Other limitations include difficulties in capturing complex scenes with multiple interacting elements and problems related to maintaining high-resolution outputs on long sequences. In general, future directions will be in the direction of integrating advanced mechanisms of attention, leveraging multi-modal data, modifying the loss function to focus more on coherence and realism even more effectively, and optimizing models for better computational efficiency. Naturally, as architectures are being proposed, improvements in grounded algorithms become possible, and training data becomes more diverse and richer, advanced text-to-video generation models can generate longer and more diverse videos that are highly realistic and exhibit temporal and spatial coherence with plausible physical interactions.

\subsection{Possible Applications and Implications}

Text-to-video generation holds transformative potential across various industries by converting textual descriptions into engaging visual content, thereby enriching communication, accessibility, and creativity. In education and e-learning, this technology facilitates a pedagogical shift, enabling teachers to explain concepts in detail through customized videos, which enhances engagement and learning~\cite{ghimire2024education, kamalov2023education}. Within the accessibility and assistive technology sectors, it offers significant assistance to individuals with disabilities, providing visual materials for the visually impaired and sign language videos for the deaf and hard-of-hearing, thereby promoting inclusiveness~\cite{trewin2018fairness}. For content creation and marketing industries, text-to-video generation serves as a cost-effective solution, allowing enterprises to produce promotional materials, personalized ads, and engaging social media content without the need for expensive, resource-intensive production processes~\cite{wu2023aigc}. In parental control and customized content applications, it enables parents to create personalized educational and entertainment videos that are safe for their children and reflect family values~\cite{viswanathan2024interaction}. Additionally, this technology can enhance the preservation and promotion of cultural heritage by transforming folkloric materials, historical accounts, and endangered languages into accessible visual narratives, thereby contributing to cultural preservation and education~\cite{jaramillo2024cultural}. In legal and forensic applications, the capability to reconstruct events and simulations from textual reports can aid investigations, court presentations, and legal education~\cite{zappala2024crime}. Text-to-video applications also support the generation of synthetic data for training and validating AI algorithms, advancing artificial intelligence development and machine learning research~\cite{anderson2022synthetic}. In the gaming industry, it facilitates the creation of sequences, such as automated cutscenes and adaptive storytelling based on player input~\cite{valevski2024realtime}. Finally, virtual reality experiences can be enhanced by automatically generating immersive environments and narratives from textual descriptions, offering personalized and interactive VR content for training, simulation, and entertainment purposes~\cite{aydin2024hrtech}. In summary, this technology is poised to impact a wide range of industry verticals by accelerating workflows, promoting inclusivity, and opening new creative horizons toward a future where content creation is continually evolving, personalized, and accessible.

\section{Conclusion}
\label{sec:conclusion}

In this survey, we reviewed the development of text-to-video (T2V) generation models, from early GAN-based and VAE-based models to current diffusion-based architectures, detailing their internal mechanisms, the limitations addressed by successive approaches, and the reasons behind major architectural shifts. We provided a review of the datasets used to train and evaluate these models, and to support reproducibility and assess the accessibility of training such models, we reported their hardware specifications, GPU counts, batch sizes, learning rates, optimizers, epochs, and other key hyperparameters. We went over the standard evaluation metrics commonly used to assess these models, presented their performance scores across these benchmarks, and discussed both the limitations of existing metrics and the emergence of more perception-aligned evaluation benchmarks. Finally, we identified the main open challenges, including alignment, long-range coherence, and computational efficiency, and outlined promising directions for future research in text-to-video generation.

\bibliographystyle{IEEEtran}
\bibliography{references}

\end{document}